\documentclass[aps,prd,reprint,superscriptaddress,nofootinbib,amsfonts,amssymb,amsmath]{revtex4-1}

\usepackage[utf8x]{inputenc}
\usepackage[dvipdf,dvips,dvipdfmx]{graphicx}
\usepackage{float}
\usepackage{wrapfig}
\usepackage{bm}
\usepackage{color}
\usepackage{comment}
\usepackage{xcolor}
\usepackage[normalem]{ulem}
\usepackage{hyperref}
\usepackage{soul}
\hypersetup{
colorlinks=true,
citecolor=blue,
citebordercolor=red,
linktoc=all,
linkcolor=blue,
urlcolor=blue
}

\catcode`\@=11
\def\lsim{\mathrel{\mathpalette\@versim<}}
\def\gsim{\mathrel{\mathpalette\@versim>}}
\def\@versim#1#2{\vcenter{\offinterlineskip
\ialign{$\m@th#1\hfil##\hfil$\crcr#2\crcr\sim\crcr } }}
\catcode`\@=12

\newcommand{\p}{\partial}

\newcommand{\al}[1]{\begin{align}#1\end{align}}
\newcommand{\bp}{\begin{pmatrix}}
\newcommand{\ep}{\end{pmatrix}}
\newcommand{\nn}{\nonumber\\}

\newcommand{\df}{\text{d}}

\newcommand{\bs}[1]{\boldsymbol}

\newcommand{\Tr}{{\rm Tr}\,}

\newcommand{\pmat}[1]{\begin{pmatrix}#1\end{pmatrix}}

\newcommand{\fn}[1]{\!\left(#1\right)}

\graphicspath{{./figs/}}

\begin{document}

\title{
Higgs scalar potential in asymptotically safe quantum gravity
}

\author{Jan M. \surname{Pawlowski}}
\affiliation{Institut f\"ur Theoretische Physik, Universit\"at Heidelberg, Philosophenweg 16, 69120 Heidelberg, Germany}
\affiliation{ExtreMe Matter Institute EMMI, GSI Helmholtzzentrum f{\" u}r
Schwerionenforschung mbH, Planckstr. 1, 64291 Darmstadt, Germany}

\author{Manuel \surname{Reichert}}
\affiliation{CP$^3$-Origins, University of Southern Denmark, Campusvej 55, 5230 Odense M, Denmark}

\author{Christof \surname{Wetterich}}
\affiliation{Institut f\"ur Theoretische Physik, Universit\"at Heidelberg, Philosophenweg 16, 69120 Heidelberg, Germany}

\author{Masatoshi \surname{Yamada}}
\affiliation{Institut f\"ur Theoretische Physik, Universit\"at Heidelberg, Philosophenweg 16, 69120 Heidelberg, Germany}

\begin{abstract}
  The effect of gravitational fluctuations on the quantum effective
  potential for scalar fields is a key ingredient for
  predictions of the mass of the Higgs boson, understanding the gauge
  hierarchy problem and a possible explanation of an---asymptotically---vanishing cosmological constant.  We find that the quartic self-interaction of the Higgs scalar field is an irrelevant coupling at
  the asymptotically safe ultraviolet fixed point of quantum
  gravity. This renders the ratio between the masses of the Higgs
  boson and top quark predictable. If the flow of couplings below the
  Planck scale is approximated by the Standard Model, this
  prediction is consistent with the observed value. The quadratic term in the
  Higgs potential is irrelevant if the strength of gravity at short
  distances exceeds a bound that is determined here as a function of
  the particle content. In this event, a tiny value of the ratio
  between the Fermi scale and the Planck scale is predicted.
\end{abstract}

\maketitle

\section{Introduction}
The quantum effective potential for the Higgs field is the central quantity for understanding the electroweak symmetry breaking in the Standard Model of particle physics (SM). The vacuum expectation value of the Higgs field $\phi_0$ is determined by the location of the minimum of the potential. It defines the Fermi scale. For given gauge couplings and Yukawa couplings it sets the mass of $W$ and $Z$ bosons as well as of quarks and charged leptons. In turn, the vacuum expectation value depends on two renormalizable couplings, the mass parameter $m_H^2$ and the quartic scalar coupling $\lambda_H$. The observable mass of Higgs boson obeys $M_H=\sqrt{2\lambda_H}|\phi_0|$.

The renormalizable couplings of the SM can be extrapolated to momenta much larger than the Fermi scale. 
In renormalization group (RG) improved perturbation theory, their running is computed with an expansion in loops. Let us now assume that the SM is part of an ``effective low-energy theory" model for scales below some transition scale $k_t$ where gravitational fluctuations decouple. Typically, $k_t$ is close to the Planck mass. In the absence of gravitational fluctuations for momenta smaller than $k_t$ all couplings are small and in the perturbative regime. For a given model the ``initial values" of $m_H^2\fn{k_t}$ and $\lambda_H\fn{k_t}$ can be extrapolated perturbatively to momenta of the order of the Fermi scale where they determine the observable quantities.

For possible predictions of the Fermi scale and the mass of the Higgs
boson, the decisive question is the predictability of $m_H^2\fn{k_t}$
and $\lambda_H\fn{k_t}$. For this issue, gravitational fluctuations
become important. For the flow of couplings at momenta larger than
$k_t$, the gravitational fluctuations strongly influence the running of
$m_H^2$ and $\lambda_H$. It has been argued that the gravitational
fluctuations drive $\lambda_H$ to a fixed-point value close to zero,
such that $\lambda_H\fn{k_t}$ has a tiny value. The extrapolation to
low momenta within the SM as effective low-energy theory has predicted~\cite{Shaposhnikov:2009pv} the mass of the Higgs
particle in accordance with later observation.

We aim here for a systematic investigation of the effects of
gravitational fluctuations on the shape of the effective scalar
potential. Beyond the Higgs sector of the SM this is relevant for
other theories with scalars, such as grand unified theories. For
cosmology, gravitational fluctuations play an important role for the
shape of scalar potentials responsible for the inflationary epoch or
dynamical dark energy. Since gravity is not perturbatively
renormalizable, any investigation of the role of gravitational
fluctuations at momentum scales larger than $k_t$ has to employ some
suitable nonperturbative method.

In the present work, we use the functional renormalization group (FRG)
for the effective average action~\cite{Wetterich:1992yh}. 
The FRG has proven to be a successful nonperturbative method for various systems in both
condensed matter and elementary particle physics. A central object
within its formulation is the scale-dependent quantum effective action
or effective average action $\Gamma_k$, which includes all effects of
quantum fluctuations with momenta larger than an IR cutoff
$k$.  The scale dependence of $\Gamma_k$ obeys an exact flow
equation~\cite{Wetterich:1992yh}.

The FRG is capable of understanding quantitatively asymptotically
safe renormalizable quantum field theories, which is crucial for
studying gravitational fluctuations near and beyond the Planck scale. For asymptotically safe
theories, the interactions do not vanish at the UV
fixed point, such that perturbative renormalizability is often not
given. Well-studied examples for nonperturbative asymptotic safety are the Wilson-Fischer fixed
point for three-dimensional scalar theories or four-dimensional
theories with four-fermion interactions, for which FRG has proven quantitative reliability. Quantum gravity presumably
belongs to this class of asymptotically safe theories.
For the quantitative study of this work, we assume asymptotic safety~\cite{Hawking:1979ig,Reuter:1996cp} as a working hypothesis, leaving the fixed-point value of the dimensionless Planck mass as a
not yet fully quantitatively determined parameter. This is sufficient
to obtain rather robust results for the effect of gravitational
fluctuations on the effective scalar potential.

The asymptotic-safety hypothesis for quantum gravity has found support
by many
investigations~\cite{Reuter:1996cp,Niedermaier:2006wt,Niedermaier:2006ns,%
  Percacci:2007sz,Reuter:2012id,Codello:2008vh,Eichhorn:2017egq,%
  Percacci:2017fkn,Eichhorn:2018yfc}. It
is crucial for this scenario that the system has a nontrivial UV
fixed point, the Reuter fixed point, at which the UV complete action is defined. Using the FRG,
the existence of such a fixed point has been investigated in pure
gravity as well as for gravity coupled to elementary
particles. The methods of approximations to the functional flow equation include
the background field
approximation~~\cite{Reuter:1996cp,Percacci:2003jz,Zanusso:2009bs,Benedetti:2009rx,Benedetti:2009gn,
Narain:2009fy,Vacca:2010mj,%
  Harst:2011zx,Eichhorn:2011pc,Eichhorn:2012va,Dona:2013qba,%
  Falls:2013bv,Falls:2014tra,Labus:2015ska,Oda:2015sma,%
  Dona:2015tnf,Eichhorn:2016esv,Gies:2016con,Eichhorn:2016vvy,%
  Hamada:2017rvn,Falls:2017lst,Eichhorn:2017als,Eichhorn:2017eht,%
  Gubitosi:2018gsl,Alkofer:2018fxj,deBrito:2018jxt,%
  Alkofer:2018baq,Falls:2018ylp},
the vertex expansion~\cite{Folkerts:2011jz,Christiansen:2012rx,Codello:2013fpa,%
  Christiansen:2014raa,Christiansen:2015rva,Meibohm:2015twa,%
  Meibohm:2016mkp,Christiansen:2016sjn,Denz:2016qks,%
  Christiansen:2017gtg,Christiansen:2017cxa,Christiansen:2017bsy,%
  Eichhorn:2018akn,Eichhorn:2018ydy},
the geometrical approach~\cite{Branchina:2003ek,Pawlowski:2003sk,Donkin:2012ud}, and the bimetric
method~\cite{Manrique:2010mq,Manrique:2010am}.  Also a gauge invariant
flow equation for quantum gravity has been proposed~\cite{Wetterich:2016ewc}.

Quantum gravity coupled to elementary particles reveals a new
predictive power for particle properties. This is connected to the
number of relevant parameters at the fixed point, which may be smaller
than the number of renormalizable couplings in the SM.
This entails that certain relations among the SM couplings
become, in principle, computable. Initial values of running
couplings become fixed at the Planck scale if they correspond to
irrelevant parameters at the fixed point. This allows a computation of
observable quantities such as the Higgs-boson mass and the top-quark
mass in the low-energy regime~\cite{Shaposhnikov:2009pv,Eichhorn:2017ylw,Eichhorn:2017lry,Eichhorn:2017muy,Eichhorn:2018whv}. The
fixed-point structure and the RG flow could
also determine the potential of scalar fields of which the time evolution characterizes the history of our Universe~\cite{Henz:2013oxa,Wetterich:2014gaa,Wetterich:2015iea,Henz:2016aoh}.
 
In this paper, we investigate quantum gravity effects on the effective scalar
potential in asymptotically safe gravity.  Taking proper account of
gauge symmetries, in our case diffeomorphism symmetry, is crucial for
quantitative reliability. For this purpose, we concentrate on a
``physical gauge fixing," which purely acts on the gauge modes among
the metric fluctuations, leaving the physical fluctuations
untouched~\cite{Wetterich:2016ewc}.  We employ the physical metric
decomposition~\cite{Wetterich:2016vxu}, where the metric fluctuations
are split into physical modes consisting of the traceless-transverse
tensor (graviton) and a scalar, and the gauge modes, which comprise a
transverse vector and a scalar. Employing the physical gauge fixing,
the two-point function becomes block diagonal in the physical and gauge modes. 
A simple relation between ghost and gauge
modes allows us to combine their contributions
to a universal measure factor, which does not depend on the value of
the scalar field~\cite{Wetterich:2016ewc}.

Our paper can also be seen as a first application of the gauge invariant flow equation employing only one macroscopic metric  field~\cite{Wetterich:2016ewc}.
In fact, at the level of truncation employed here, there is no difference between the background formalism with physical gauge fixing and the gauge invariant flow equation.
The proposed universal measure contribution~\cite{Wetterich:2016ewc} comes out directly in our approximation for the background formalism.
The flow equation for the effective potential is the same for the truncated background formalism and the gauge invariant flow equation.
The contribution to the gauge invariant flow equation from physical fluctuations involves formally nonlocal projections.
This projection is implicitly performed in the background field formalism by the inversion of the second functional derivative of the effective action in presence of the physical gauge fixing term.
The relevant projected differential operators for the graviton and the physical scalar metric fluctuations are second-order differential operators.
No nonlocality is encountered explicitly.

The propagator and interactions for the physical modes are derived here from a gauge invariant effective action.
In the background formalism, this is an approximation, while for the gauge invariant flow equation, this is a genuine property.
We also compute the flow of the mass term and quartic coupling by taking derivatives of the flow equation for the effective potential. 
In the truncated background formalism, this is an approximation, while for the gauge invariant flow equation, this is an exact property.
Here we do not enter into the discussion of whether the employed one-loop form of the gauge invariant flow equation is itself an approximation, whether it can be made exact by a suitable definition of the macroscopic field~\cite{Wetterich:2016ewc}.

\begin{figure}
\includegraphics[width=\linewidth]{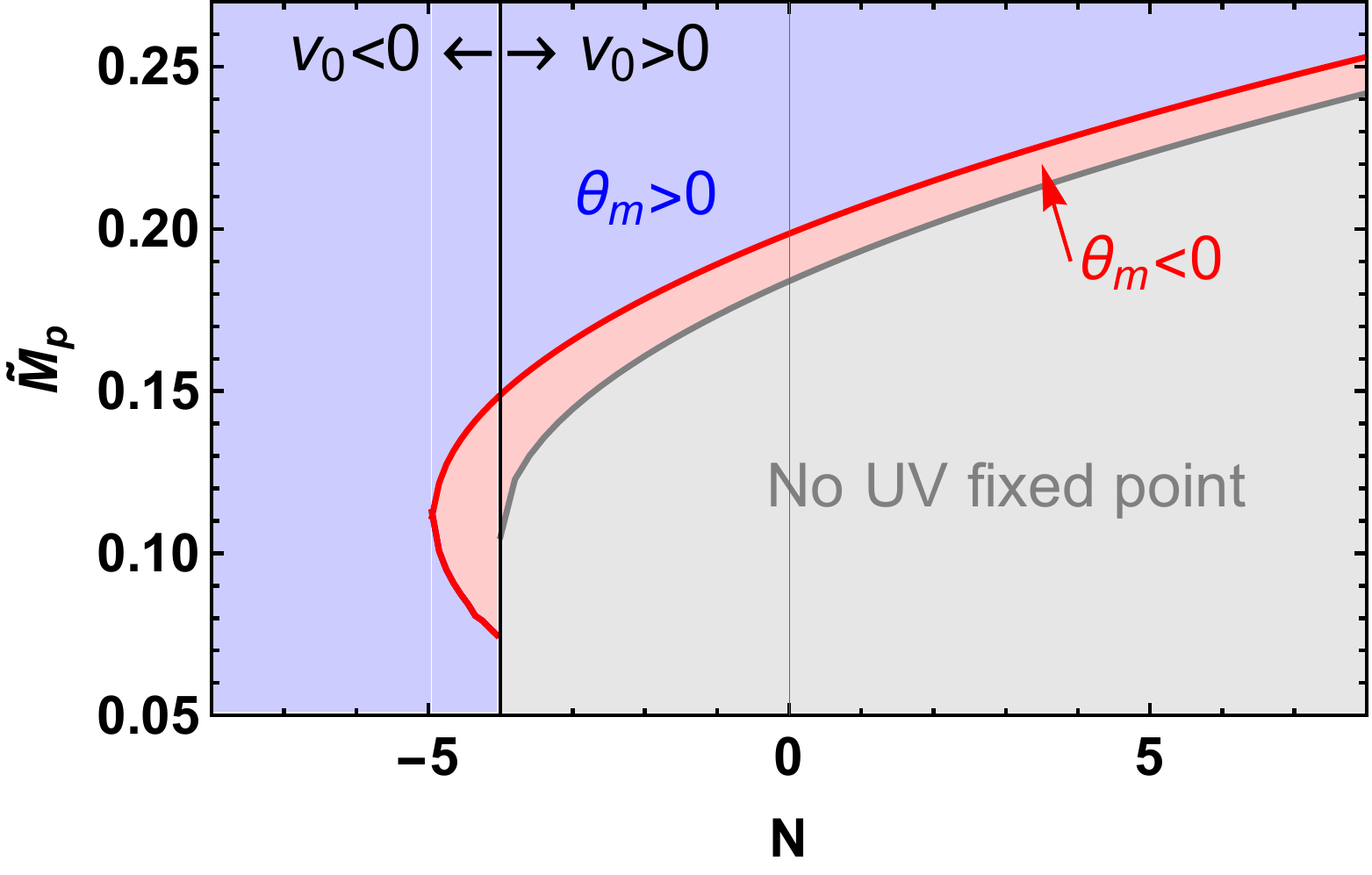}
\caption{Existence of the UV fixed point and sign of the critical
  exponent of the scalar mass term, as a function of the fixed-point value
  $\tilde M_\text{p}$ for the running dimensionless Planck mass and
  the number of effective particle degrees of freedom
  $N=N_S+2N_V-2N_F$. The grey line is the critical value of the Planck
  mass $\tilde M_\text{p,c}$. In the red-colored region, the critical
  exponent of the scalar mass term becomes negative, i.e., this coupling is
  irrelevant, while it is relevant in the blue-colored region.  The
  quartic scalar coupling is irrelevant whenever a fixed point exists.
}
\label{fig:scalar number and Planck mass} 
\end{figure}

Our main results for the effects of gravitational fluctuations on the
scalar effective potential are the following: (i) A UV fixed point for
the cosmological constant (value of scalar potential at its minimum)
exists provided the dimensionless squared Planck mass
$\tilde M_\text{p}^2=M_\text{p}^2/k^2$ is above a minimal value
$\tilde M_\text{p,c}^2$. This value depends on the number $N$ of
particle degrees of freedom, as shown in Fig.\,\ref{fig:scalar number
  and Planck mass}. (ii) The quartic scalar coupling $\lambda_H$ of
the Higgs boson is an irrelevant coupling at the UV fixed point. For
large $k^2$, the gravitational fluctuations drive it very close to
zero, enforcing an initial value for the low energy effective
theory $\lambda_H\fn{k_t}\approx 0$. For a given
low-energy theory at momentum scales below the Planck mass and a given
observed mass of the top quark, the mass of the Higgs boson $M_H$
becomes predictable. If the low-energy theory is the SM, the predicted
value~\cite{Shaposhnikov:2009pv} is $M_H=126$\,GeV with a few giga-electron-volts uncertainty. For a higher-loop computation and dependence on the top mass, see~Refs.\,\cite{Bezrukov:2012sa,Buttazzo:2013uya}. 
(iii) For small enough $\tilde M_\text{p}^2$ (red region
in Fig.\,\ref{fig:scalar number and Planck mass}), the quadratic term
in the scalar potential is also an irrelevant parameter. Then, the model
is predicted to be located on the critical phase transition surface of
the vacuum electroweak phase-transition, realizing self-organized
criticality. The gauge hierarchy~\cite{Gildener:1976ai,Weinberg:1978ym} of a tiny ratio
between the Fermi scale and Planck scale could then be explained by the
resurgence mechanism~\cite{Wetterich:2016uxm}.

First indications that a quartic scalar coupling could be an irrelevant parameter can be found in Refs.\,\,\cite{Percacci:2003jz,Zanusso:2009bs,Narain:2009fy,Rodigast:2009zj,Mackay:2009cf,Vacca:2010mj}.
At the time of the prediction \cite{Shaposhnikov:2009pv} of the Higgs-boson mass, important uncertainties about the sign and magnitude of the anomalous dimension for the quartic coupling of the Higgs scalar persisted, however.
Emphasis on the dominant role of the graviton fluctuations (``graviton approximation'') has shown \cite{Wetterich:2017ixo} the positive sign of the anomalous dimension $A$ and estimated its magnitude to be of the order 1.
These are precisely the requirements for the prediction of the mass of the Higgs boson~\cite{Shaposhnikov:2009pv}.
In the present paper, we confirm the graviton domination by an explicit computation of the contribution of all other fluctuations, including the universal measure term for a physical gauge fixing.
This allows for a quantitative comparison with the dominant graviton contribution.
We also show that extensions of the truncation do not alter the main conclusion that quartic scalar couplings are irrelevant parameters at the UV fixed point for asymptotic safety.

The quantitative precision of the present approach allows for the first time an estimate under which circumstances the scalar mass term can be an irrelevant coupling~\cite{Wetterich:2016uxm}. 
Typically, this occurs for $A>2$.
The size of $A$ depends strongly on the fixed-point value of the dimensionless Planck mass.
This value depends, in turn, on the precise particle content of the model and requires a computation of the flow equation with similar detail as the one for the effective potential investigated here.
Only once this task is accomplished, a definite statement on the predictive power of asymptotic safety for the gauge hierarchy will be possible.

This paper is organized as follows. In the next section, we present
the flow equation for the scalar potential. The technical aspects are given in Appendix~\ref{formulations}.  In
Sec.\,\ref{section of Fixed point and critical exponent of scalar
  potential}, we analyze the fixed-point structure and the critical
exponent for the cosmological constant.  Section\,\ref{scalar
  interaction part} investigates the critical exponents for the
scalar mass term and the quartic interaction of the scalar field. We
address here the predictive power of quantum gravity for properties of
the Higgs scalar.
 Section\,\ref{Robustness check} discusses the robustness of our results by extending the truncation and varying the cutoff function.
 Section\,\ref{discussion section} is devoted to
summarizing our results and to discussing their robustness and remaining
quantitative uncertainty.

\section{Flow of the scalar potential}
The flow equation for the effective scalar potential is extracted from
the exact flow equation for the effective average action by
taking space- and time-independent field values for the scalar field
configuration.  The flow is evaluated for a flat spacetime geometry
that we take here to be Euclidean.  The crucial quantity for the flow
equation is the inverse propagator, which is given by the matrix of
second functional derivatives $\Gamma_k^{(2)}$ of the effective
action, evaluated for the given scalar and metric
fluctuation. Precision and robustness of results depend on the
validity of the approximations used for $\Gamma_k^{(2)}$.  Gravity is
a local gauge theory, with gauge transformations associated to
diffeomorphism or general coordinate transformations. It is a crucial
issue to take the gauge symmetry properly into account. The
fluctuations around any given metric configuration can be split into
gauge fluctuations or gauge modes, and physical fluctuations or
modes. 
The gauge modes correspond to the infinitesimal changes of the given metric induced by an infinitesimal gauge transformation.

We follow the standard treatment of functional integrals for
gauge theories, implementing gauge fixing and the associated Faddeev-Popov determinant.
We impose a particular physical gauge fixing~\cite{Wetterich:2016ewc}.  A physical gauge fixing acts only
on the gauge modes. 
Choosing a decomposition of the
metric fluctuations into physical modes and gauge
modes~\cite{Wetterich:2016vxu}, the physical gauge-fixing term renders $\Gamma_k^{(2)}$ effectively
block diagonal, with separate blocks for the physical modes and the
gauge modes.  Furthermore, imposing the physical gauge constraint on
the fields in the effective action leaves a gauge invariant effective
action~\cite{Wetterich:2016ewc}. We can therefore employ an ansatz
where the effective action consists of a gauge invariant part
$\bar \Gamma_k$ plus a gauge-fixing part.  For the inverse propagator
$\Gamma_k^{(2)}$, the block for the physical modes is given by the
second functional derivative of $\bar \Gamma_k$. This is an
important advantage, since gauge symmetry severely restricts the form
of $\bar \Gamma_k$.

We find that the contribution of the gauge modes, together with the
contribution from the Faddeev-Popov determinant or the corresponding
ghosts, results in a simple universal contribution to the flow
equation. This ``measure contribution" depends on the metric but not on
the values of scalar fields. For the flow of the effective potential,
it only concerns an overall constant but not the field dependence.

What remains to be done is an effective approximation for the physical
inverse propagator $\bar \Gamma_k^{(2)}$. This is done by making an
ansatz for the gauge invariant effective action $\bar \Gamma_k$. We
approximate the gravitational part of $\bar \Gamma_k$ by the
Einstein-Hilbert action, with coefficient of the curvature scalar
given by the running or scale-dependent squared Planck mass
$M_\text{p}^2\fn{k}$. The cosmological constant is included as part of
the effective scalar potential, namely, its value at the minimum. We
discuss in the conclusions how this ansatz can also incorporate
effects of higher-derivative invariants in $\bar \Gamma_k$, as $R^2$
or $R_{\mu\nu}R^{\mu\nu}$. This is done by an adaptation of the
definition of $M_\text{p}^2\fn{k}$.

According to our assumption of asymptotic safety the running Planck
mass has to scale at and near the UV fixed point proportional to
$k$, 
\al{
M_\text{p}^2\fn{k}=\tilde M_{\text{p}*}^2\,k^2 \,.
}
The fixed-point value $\tilde M_{\text{p}*}^2$ depends on the particular
model. For the purpose of this paper we treat it as an unknown
parameter. Some of the predictions depend on the precise value of this
parameter, while others such as the quartic coupling $\lambda_H$ being
an irrelevant parameter are independent of the precise value.

We first consider a single real scalar field $\phi$ coupled to
gravity. The detailed steps of the computation along the lines
sketched above are displayed in Appendix~\ref{formulations}.  We
obtain for the flow of the effective potential $U\fn{\rho}$ at fixed
$\rho=\phi^2/2$ a differential equation with a rather simple form,
\al{ \p_t U=k \p_k U=\tilde \pi_2+\tilde \pi_0 +\tilde \eta\,.
\label{flow equation of U}
}
Here, $\tilde \pi_2$ is the contribution of the graviton fluctuations
corresponding to the traceless-transverse metric fluctuations; the
term $\tilde \pi_0$ combines the physical scalar fluctuations, both
from $\phi$ and the physical scalar mode in the metric
fluctuations. Finally, $\tilde \eta$ is the measure
contribution. Employing a Litim-type cutoff
function~\cite{Litim:2001up}, the terms are given by
\al{
  \tilde \pi_2&=\frac{5}{24\pi^2}\left(1 -\frac{\eta_g}{8}\right)
  \frac{k^4}{1- v} \,,\nonumber\\[2ex]
  \tilde \pi_0&=\frac{1}{24\pi^4}\bigg[ \left(1-\frac{\eta_g}{8}
    \right)\left(1+{\tilde U'+2\tilde \rho \tilde U''}\right) \nn
    &\qquad+     \frac{3}{4}\left(1-\frac{\eta_\phi}{6}\right) \left(1-
      \frac{v}{4}\right)  \bigg] 
\label{pi0 contribution}\\
  &\quad \times     \frac{k^4}{\left[ \left(1-
        v/4 \right) \left(1+\tilde U'+2\tilde \rho \tilde U''
      \right) +3\tilde \rho \tilde U'{}^2/\tilde M_\text{p}^2\right]}\,,\nn[2ex]
\tilde \eta&=-\frac{k^4}{8\pi^2}\, .\nonumber
}
Here, we have defined the dimensionless quantities, 
\al{
&{\tilde U}\fn{\tilde \rho}=U\fn{\rho}/k^4 \,,&
&\tilde \rho=Z_\phi\rho/k^2 \,,&
}
with $Z_\phi$ the coefficient of the scalar kinetic term, and primes denoting derivatives with respect to $\tilde \rho$.
The dimensionless ratio
\al{
v\fn{\rho}=\frac{2U\fn{\rho}}{M_\text{p}^2k^2}=\frac{2\tilde U\fn{\rho}}{\tilde M_\text{p}^2}\,,
}
depends on $\rho$. The poles at $v=1$ and $v=4$ correspond to tachyonic instabilities in the graviton and the scalar mode of metric fluctuation propagator, respectively. They are not reached by the flow.
We furthermore define $\eta_g=-\p_t  \ln {\tilde M}_\text{p}^2=2-\p_t \ln {M}_\text{p}^2$ and the anomalous dimension of the scalar field, $\eta_\phi=-\p_t \ln Z_\phi$.

We will see that at the fixed point the minimum of $U\fn{\rho}$ occurs for $\rho=0$. In the vicinity of this point we can neglect the term $3\tilde \rho \tilde U^2/\tilde M_\text{p}^2$ in the denominator of \eqref{pi0 contribution}, such that also the effect of scalar fluctuations becomes block diagonal,
\al{
\tilde \pi_0 &\simeq \tilde\pi_{0,g}+\tilde\pi_{0,\phi}\,,
\label{approximation for the spin 0 modes}
}
with
\al{
\tilde\pi_{0,g}&=\frac{1}{24\pi^2}\left(1 -\frac{\eta_g}{8}\right)\frac{k^4}{1- v/4}\,,
\nn
\tilde\pi_{0,\phi}&=\frac{1}{32\pi^2}\left( 1-\frac{\eta_\phi}{6}\right) \frac{k^4}{1+\tilde U'+2\tilde\rho\tilde U''}\,.
\label{contribution from pi0}
}
The first contribution $\tilde \pi_{0,g}$ arises from the scalar mode in the metric, while the second term $\tilde \pi_{0,\phi}$ is the standard flow contribution from scalar fields in flat space~\cite{Wetterich:1992yh}.

The flow equation \eqref{flow equation of U} holds at constant $\rho$. For a discussion of a fixed point and the behavior close to the fixed point, we have to translate to the flow at constant $\tilde \rho$. Furthermore, we are interested in the flow of the dimensionless scalar potential $\tilde U\fn{\tilde \rho}$. Generalizing to $N$ real scalars with O($N$) symmetry and employing the approximation \eqref{approximation for the spin 0 modes}, we obtain the beta function or flow generator for $\tilde U$ as
\al{
\p_t \tilde U&=-4\tilde U +(2+\eta_\phi)\tilde \rho\tilde U' \nn
&\quad+\frac{1}{24\pi^2}\left(1 -\frac{\eta_g}{8}\right)\left[ \frac{5}{1- v}+\frac{1}{1- v/4}\right]
+\frac{\Delta N-4}{32\pi^2}\nn
&\quad
+\frac{1}{32\pi^2}\left( 1-\frac{\eta_\phi}{6}\right)\left[ \frac{1}{1+ \tilde U'+2\tilde \rho \tilde U''} +\frac{N-1}{1+\tilde U'} \right]\,.
\label{beta function of effective scalar potential}
}
The dependence of the gravitational contributions on $\tilde \rho$ arises through the quantity $v=v\fn{\tilde \rho}$. Also, the dimensionless Planck mass enters in \eqref{beta function of effective scalar potential} only through $v$.
The first two terms on the right-hand side of \eqref{beta function of effective scalar potential} are the canonical scaling of the effective potential.

We have extended the scalar sector to $N$ scalars with SO($N$) symmetry.
For scalar theories with SO($N$) symmetry, the term proportional to $N-1$ arises from the fluctuations in the Goldstone directions.
For the Higgs doublet, one has $N=4$.
Furthermore, we have included in \eqref{beta function of effective scalar potential} the contribution of fluctuations beyond the gravitational degrees of freedom and the Higgs sector. For massless particles, as gauge bosons or chiral fermions, they contribute to $\p_t U$ a field-independent term $\Delta N/(32\pi^2)$, where
\al{
\Delta N=\Delta N_S+2N_V-2N_F\,,
\label{number of particles}
}
with $\Delta N_S$ the number of additional scalars, $N_V$ the number of gauge bosons (with two physical degrees of freedom each), and $N_F$ the number of Weyl fermions. For the SM-matter content, this number is $\Delta N_\text{SM}=-66$; for a grand unified gauge theory based on SO(10), one has $\Delta N_\text{GUT}=N_S+10$ with $N_S$ the total number of real scalars beyond a complex 10-representation.

Around the origin at $\tilde \rho=0$, we expand
\al{
\tilde U=\tilde V +\tilde m_H^2\tilde \rho+\frac{\tilde \lambda_H}{2}\tilde \rho^2+\cdots.
}
Inserting into \eqref{beta function of effective scalar potential} yields the beta function for each coupling,
\begin{widetext}
\al{
\p_t {\tilde V}&=-4{\tilde V} +\frac{1}{24\pi^2}\left(1 -\frac{\eta_g}{8}\right)\left[ \frac{5}{1- v_0}+\frac{1}{1- v_0/4}\right]
+\frac{N}{32\pi^2}\left( 1-\frac{\eta_\phi}{6}\right)\frac{1}{1+\tilde m_H^2}
+\frac{\Delta N-4}{32\pi^2}\,,
\label{beta function of cosmological constant}
\\[1ex]
\p_t {\tilde m_H^2}&=(-2+\eta_\phi){\tilde m_H^2} +\frac{{\tilde m_H^2}}{48\pi^2\tilde M_\text{p}^2}\left(1 -\frac{\eta_g}{8}\right)\left[ \frac{20}{(1- v_0)^2}+\frac{1}{\left(1- v_0/4\right)^2}\right] 
-\frac{(N+2)\tilde \lambda_H}{32\pi^2}\left( 1-\frac{\eta_\phi}{6}\right) \frac{1}{(1+\tilde m_H^2)^2}\,,
\label{beta function of scalar mass}
\\[1ex]
\p_t {\tilde \lambda_H}&=2\eta_\phi{\tilde \lambda_H} +\frac{\tilde \lambda_H}{48\pi^2\tilde M_\text{p}^2}\left(1 -\frac{\eta_g}{8}\right)\left[ \frac{20}{(1- v_0)^2}+\frac{1}{\left(1- v_0/4\right)^2}\right]\nn
&\quad
+\frac{\tilde m_H^4}{48\pi^2\tilde M_\text{p}^4}\left( 1-\frac{\eta_g}{8}\right)\left[ \frac{80}{(1-v_0)^3}+\frac{1}{(1-v_0/4)^3}\right]
+\frac{(N+8)\tilde \lambda_H^2}{16\pi^2}\left( 1-\frac{\eta_\phi}{6}\right) \frac{1}{(1+\tilde m_H^2)^3}\,.
\label{beta function of quartic coupling}
}
\end{widetext}
Here, we have defined the dimensionless renormalized parameters as $\tilde V=U\fn{\rho=0}/k^4$, $\tilde m_H^2=m_H^2/(Z_\phi k^2)$, $\tilde \lambda_H=\lambda_H/Z_\phi^2$, and $v_0=2\tilde V/\tilde M_\text{p}^2$. In general, Eqs.\,\eqref{beta function of cosmological constant}--\eqref{beta function of quartic coupling} receive contributions from the term $3\tilde \rho \tilde U'{}^2/\tilde M_\text{p}^2$ in \eqref{pi0 contribution}, neglected in \eqref{contribution from pi0} and \eqref{beta function of effective scalar potential}. These contributions are proportional to higher orders of the coupling constants, e.g., $\tilde m_H^4$ or $\tilde m_H^2\tilde \lambda_H$. If the fixed point of matter interactions occurs for $\tilde m_{H*}^2=\tilde \lambda_{H*}=0$, these terms do not contribute to the critical exponents defined below; see \eqref{linear RG equation}.

The effects of Yukawa couplings or gauge couplings to the Higgs sector correspond to the standard perturbative contributions to the beta functions. These effects are small and are not included in our discussion of the UV fixed point. In the present approximation, the additional particles only influence the flow of $\tilde V$, with no direct influence on \eqref{beta function of scalar mass} and \eqref{beta function of quartic coupling}.

We observe that for $\tilde m_H^2=0$, as appropriate for the UV fixed point, and $\eta_\phi=0$ the fluctuations of the Higgs scalar ($N=4$) cancel the measure contribution, $\tilde \pi_{0,\phi}+\tilde \eta=0$. The two last terms in \eqref{beta function of cosmological constant} can then be collected into $\Delta N/(32\pi^2)$. 

Instead of the cosmological constant $\tilde V$, it is useful to introduce the beta function of the dimensionless quantity $v_0=2\tilde V/\tilde M_\text{p}^2$, which reads
\al{
\p_t v_0&=(-4+\eta_g)v_0\nn
&\quad
+\frac{1}{12\pi^2{\tilde M}_\text{p}^2}\left(1 -\frac{\eta_g}{8}\right)\left[ \frac{5}{1- v_0}+\frac{1}{1- v_0/4}\right] \nn
&\quad
+\frac{N}{16\pi^2{\tilde M}_\text{p}^2}\left( 1-\frac{\eta_\phi}{6}\right)\frac{1}{1+\tilde m_H^2}
+\frac{\Delta N-4}{16\pi^2{\tilde M}_\text{p}^2}\,.
\label{cosmological constant in terms of v}
} 
Equations.\,\eqref{beta function of scalar mass}--\eqref{cosmological
  constant in terms of v} constitute a system of three coupled
nonlinear differential equations. They are solved numerically. We
employ $\eta_g=0$ as appropriate for the UV fixed point and also
neglect the presumably small scalar anomalous dimension $\eta_\phi$,
which arises from the flow of the kinetic term for $\phi$. The result
of the numerical solution is shown in Fig.\,\ref{fig:scalar number and
  Planck mass}. For the purpose of this figure, we define an effective
$N=\Delta N+4$. For $\tilde M_{\text{p}*}$ outside the grey region, we
find indeed a UV fixed point. For $N\geq -4$, our assumption of
asymptotic safety holds only if gravity is not too strong, such that
$\tilde M_{\text{p}*}$ remains above the lower bound indicated by the
grey line in Fig.\,\ref{fig:scalar number and Planck mass}.

\section{Fixed point and critical exponents for the scalar potential}
\label{section of Fixed point and critical exponent of scalar potential}
Let us now investigate the fixed-point structure and the critical
exponents for the scalar potential $U$.  To this end, we need the
dimensionless Planck mass $\tilde M_\text{p}$, which enters directly
in \eqref{beta function of scalar mass}, and \eqref{beta function of
  quartic coupling} and indirectly through
$v_0=2\tilde V/\tilde M_\text{p}^2$. In order to close the system, the
beta function for $\tilde M_\text{p}$ would be needed.  The latter
depends on the particle content of the theory. Its computation is also
influenced by the truncations of the system, the choices of gauge
parameters, and the regulator. We assume here only that a fixed point
of the Planck mass exists and treat $\tilde M_\text{p}$ as a free
constant parameter. Since the Newton coupling is defined as
$G_\text{N}=1/(8\pi M_\text{p}^2)$, a small value of the Planck mass
corresponds to a strong interaction of gravity.  A constant
$\tilde M_\text{p}$ results in $\eta_g=0$.

We also assume that the system has a Gaussian-matter fixed point,
namely, that a nontrivial fixed point is present in the gravity
sector, while gauge and Yukawa couplings in the matter sector vanish
at the fixed point. We will discuss in Sec.\,\ref{discussion section}
the possibility that the matter interactions have a nontrivial fixed
point and their effects on the critical exponents. We neglect the
small~\cite{Eichhorn:2017als} anomalous dimension $\eta_\phi$. With
these assumptions, we obtain the fixed points and the critical
exponents as functions of $\tilde M_\text{p}$.

\subsection{Critical exponents}
Before discussing the structure of the beta functions, we briefly recall the definition of the critical exponents.
We denote the renormalized couplings that span the theory space by $g=\{g_1,...,g_i,...\}$.
The RG equations are generally given by
\al{
\p_t {\tilde g}_i = \beta_i\fn{\tilde g}=-d_i{\tilde g}_i +f_i\fn{\tilde g}\,,
\label{general RG equations}
} where $\tilde g_i=g_i k^{-d_i}$ is a dimensionless coupling and
$d_i$ is the mass dimension of $g_i$.  The first term on the
right-hand side reflects the canonical scaling, whereas the second one
is the fluctuation contribution obtained from the flow equation.  Suppose that there exists a nontrivial fixed
point $\tilde g_{i*}$. 
The critical exponents characterize the RG flow in the vicinity of the fixed point. We therefore linearly expand
the RG equation \eqref{general RG equations} \al{ \p_t {\tilde g}_i =
  \sum_j\frac{\p \beta_i}{\p \tilde g_j}\bigg|_{\tilde g=\tilde
    g_*}(\tilde g_j-\tilde g_{j*})=-T_{ij}(\tilde g_j-\tilde
  g_{j*})\,.
\label{linear RG equation}
}
The matrix $T$ is the stability matrix, and its eigenvalues, denoted by $\theta_l$, are the critical exponents.
The solution to \eqref{linear RG equation} is 
\al{
\tilde g_i= \tilde g_{i*} + \sum_l C_l V_i^l\left(\frac{k}{\mu}\right)^{-\theta_l}\,,
\label{general RG flow}
}
where $V_i^l$ is the matrix that diagonalizes the stability matrix and $C_l$ are constant coefficients given at a reference scale $\mu$.
Positive critical exponents correspond to relevant couplings, whereas the irrelevant couplings have negative critical exponent.
As $k$ is lowered, the irrelevant couplings flow toward their fixed-point values. Defining a theory at some UV fixed point, the irrelevant couplings take their fixed-point values, setting $C_l=0$ for all $l$ with $\theta_l<0$ in \eqref{general RG flow}. The coefficients $C_l$ for the relevant parameters are the only free parameters of the theory.

Close to a fixed point with $\tilde m_{H*}^2=0$, $\tilde \lambda_{H*}=0$, we linearize in $\tilde m_H^2$ and $\tilde \lambda_H$.
Taking $N=4$, the flow equations simplify to
\al{
\p_t v_0&=-4v_0
+\frac{\Delta N}{16\pi^2 \tilde M_\text{p}^2}\nn
&\quad+\frac{1}{12\pi^2 \tilde M_\text{p}^2}\left[ \frac{5}{1-v_0}+\frac{1}{1-v_0/4}\right]
-\frac{\tilde m_H^2}{4\pi^2 \tilde M_\text{p}^2}\,,
\label{simplified v equation}
\\
\p_t \tilde m_H^2&=-2\tilde m_H^2  -\frac{3\tilde \lambda_H}{16\pi^2}\nn
&\quad
+\frac{\tilde m_H^2}{48\pi^2 \tilde M_\text{p}^2}\left[ \frac{20}{(1-v_0)^2}+\frac{1}{(1-v_0/4)^2}\right]\,,
\label{RG equation of scalar mass}\\
\p_t\tilde \lambda_H&=\frac{\tilde\lambda_H}{48\pi^2 \tilde M_\text{p}^2}\left[ \frac{20}{(1-v_0)^2}+\frac{1}{(1-v_0/4)^2}\right]\,.
\label{RG equation of quartic coupling}
} The stability matrix in the space of couplings $v_0$,
$\tilde m_H^2$, $\tilde \lambda_H$ follows by taking derivatives at
the fixed-point values $v_{0*}$,
$\tilde m_{H*}^2=\tilde \lambda_{H*}=0$, \al{ T=\pmat{
    4-A & \displaystyle \frac{1}{4\pi^2 \tilde M_\text{p}^2} & 0\\
    0 & 2-A & \displaystyle \frac{3}{16\pi^2}\\[10pt]
    0 & 0 & -A }\,.
\label{stability matrix T}
}
The quantity
\al{
A=\frac{1}{48\pi^2 \tilde M_\text{p}^2}\left[ \frac{20}{(1-v_0)^2}+\frac{1}{(1-v_0/4)^2}\right]\,,
\label{anomalous dimension A}
} depends only on $\tilde M_\text{p}^2$ and $v_0$, not on $\Delta
N$. The eigenvalues of $T$ are simply the diagonal elements of the
matrix \eqref{stability matrix T}. In a more complete setting with a
beta function for $\tilde M_\text{p}^2$ depending on $v_0$ and
$\tilde m_H^2$, the extended stability matrix involves mixing effects
with the sector describing the flow of $\tilde M_\text{p}^2$, as well
as possibly with other flowing parameters in the gravitational
sector. These mixing effects are neglected in the present work. They
concern only the critical exponent for $v_0$. Since the beta functions
for $\tilde m_H^2$ and $\tilde \lambda_H$ vanish for $\tilde m_H^2=0$,
$\tilde \lambda_H=0$, the derivative of these functions with respect
to $\tilde M_\text{p}^2$ does not contribute at the fixed point. As a
consequence, the critical exponents for $\tilde m_H^2$ and
$\tilde \lambda_H$ are not affected by the mixing and remain to be
given by $2-A$ and $-A$.

These features allow for rather robust predictions of the critical
exponents once the fixed-point values for $\tilde M_\text{p}^2$ and
$v_0$ are known. One only needs the computation of $A$ in
\eqref{anomalous dimension A}. The first term in \eqref{anomalous
  dimension A} is the graviton contribution from $\tilde \pi_2$. It
typically exceeds the second term by more than a factor $20$. This
validates the graviton approximation for the computation of the
critical exponents in Ref.\,\cite{Wetterich:2016ewc}.

\subsection{Cosmological constant}
The value of the cosmological constant $v_0$ near the fixed point has
a substantial influence on the size of the gravitational
fluctuations. We determine here the fixed-point value and the
associated critical exponent.
\subsubsection{Fixed point as a function of the Planck mass}
We first look for a possible fixed point of the cosmological constant
as a function of the Planck mass by setting the right-hand side of
\eqref{simplified v equation} to zero. We concentrate on $N=4$, as
appropriate for the Higgs doublet.  The resulting quadratic equation
for $v_{0*}$ admits two solutions: One is a UV fixed point, and the other
is an IR one. For instance, in the graviton approximation, namely,
taking only the contribution of the traceless transverse mode $\pi_2$
into account, one finds for $\Delta N=0$~\cite{Wetterich:2017ixo} 
\al{
  v_{0*}^{(\text{UV})}&=\frac{1}{2}\left( 1-\sqrt{1-\frac{5}{12\pi^2\tilde M_{\text{p}*}^2}}\right)\,,\\
  v_{0*}^{(\text{IR})}&=\frac{1}{2}\left(
    1+\sqrt{1-\frac{5}{12\pi^2\tilde M_{\text{p}*}^2}}\right)\,.  }
These values obtain small corrections from the scalar mode in the
metric.  In Fig.\,\ref{fig:fixed point value of cosmological
  constant}, we plot the dependence of the fixed-point value of
$v_{0}$ on the Planck mass for $N=4$ and $\Delta N=0$. We show both
the full result (blue) and the graviton approximation (only $\pi_2$,
red). We conclude that the graviton approximation is valid up to
corrections approximately $10\%$.

\begin{figure}
\includegraphics[width=\linewidth]{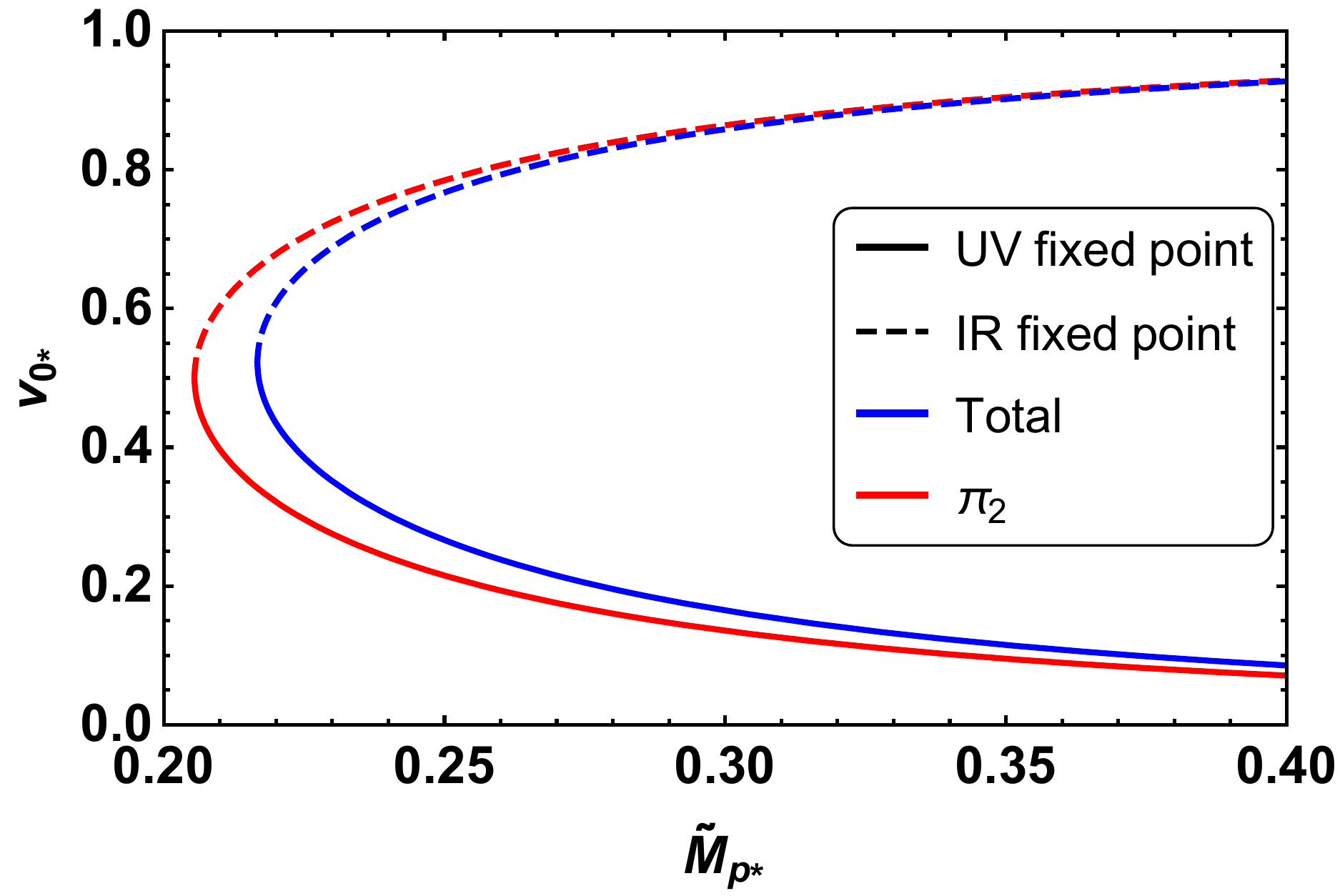}
\caption{The fixed-point value of the cosmological constant $v_{0*}$
  as a function of ${\tilde M}_{\text{p}*}$ for $N=4$ and
  $\Delta N=0$. The effects of the graviton fluctuations dominate, as
  shown by including only $\pi_2$ in \eqref{flow equation of U}.  }
\label{fig:fixed point value of cosmological constant} 
\end{figure}
In the limit $\tilde M_{\text{p}*}\to \infty$, the UV fixed point
$v_{0*}^\text{(UV)}$ merges to the Gaussian fixed point, while the IR
one converges to 1.  In contrast, for decreasing
$\tilde M_{\text{p}*}$, these two fixed points approach each other.
They merge at a critical value $\tilde M_\text{p,c}=0.217$.  For
$\tilde M_\text{p}<\tilde M_\text{p,c}$, no fixed point exists, and
quantum gravity is not asymptotically safe. The critical value
$\tilde M_\text{p,c}$ depends on the particle content in the UV.  We
plot $\tilde M_\text{p,c}$ as a function of $N$ in
Fig.\,\ref{fig:scalar number and Planck mass}.  Here, $N$ stands for
$4+\Delta N$. We include the range of negative $N$, as relevant for a
sufficient number of fermions; see \eqref{number of
  particles}. 
  Indeed, gauge bosons and scalars give a positive
contribution to $N$, while for fermions, the contribution is
negative. The value $\tilde M_\text{p,c}^2$ is obtained by imposing at
$v_0$ the simultaneous requirements $\p_t v_0=\beta_v=0$ and
$\p \beta_v/\p v_0=0$. For a given $N$, asymptotic safety is only
realized for $\tilde M_\text{p}>\tilde M_\text{p,c}\fn{N}$, as is visible
in Fig.\,\ref{fig:scalar number and Planck mass}.  According to
\eqref{stability matrix T} and \eqref{anomalous dimension A}, the value
$\tilde M_\text{p,c}$ corresponds to $A=4$. In the vicinity of this
critical value, both the mass term and the quartic scalar coupling are
irrelevant couplings, with critical exponents $-2$ and $-4$,
respectively.

The IR fixed point for $v_0$ always exists for a given
$\tilde M_{\text{p}*}>\tilde M_\text{p,c}$, since the flow of $v_0$ is
always stopped before the pole of the beta function at $v_0=1$ is
reached. On the other hand, $\tilde M_\text{p}$ depends on $v_0$. The
function $\tilde M_{\text{p}*}\fn{v_0}$ corresponds to a curve in
Fig.\,\ref{fig:fixed point value of cosmological constant} that is not
computed in the present work. The UV fixed point corresponds to the
intersection of this curve with the curve shown in
Fig.\,\ref{fig:fixed point value of cosmological constant}. If there
exist two intersection points, both with the solid and the dashed
lines, both a UV and an IR fixed point exist.  If present, one could
alternatively define the theory at the IR fixed point. Since
$\p \beta_v/\p v_0>0$ at the IR fixed point, one infers from
\eqref{stability matrix T} and \eqref{anomalous dimension A} that
$A>4$. Thus, all three parameters $v_0$, $\tilde m_H^2$, and
$\tilde \lambda_H$ are irrelevant couplings.  In the present paper, we
do not pursue this possible alternative and rather concentrate on the
UV fixed point.

Once $\tilde M_\text{p}^2$ increases as this coupling moves away from the UV fixed point, the IR fixed point for $v_0$ approaches the pole in the beta function. Indeed, the value $v_{0}=1$ corresponds to a pole of the propagator of the graviton. The existence of the IR fixed point close to $v_0=1$ induces strong fluctuation effects of the graviton in the IR regime and could be a key point to resolve the cosmological constant problem~\cite{Wetterich:2017ixo,Wetterich:2018poo}.

For general $\Delta N$, we parametrize the ratio between the scalar and tensor gravitational contributions by 
\al{
w_\text{s}\fn{v}=\frac{\tilde \pi_{0,g}}{\tilde \pi_2}=\frac{1-v}{5(1-v/4)}\,.
}
The fixed points for $v_0$ occur for 
\al{
v_{0*}=\frac{1}{2}\left( 1+z\Delta N \pm \sqrt{(1-z\Delta N)^2-\frac{80z(1+w_\text{s})}{3}}\right)\,,
}
with 
\al{
z=\frac{1}{64\pi^2\tilde M_\text{p}^2}\,.
}
For large negative $\Delta N$, the UV fixed point occurs for a negative value, 
\al{
v^\text{(UV)}_{0*}\approx z\Delta N +\frac{20z(1+w_\text{s})}{3(1-z\Delta N)}\,.
\label{UV v approx}
}
As long as $z\Delta N$ remains small as compared to 1, one has $w_\text{s}\approx 1/5$ and
\al{
v^\text{(UV)}_{0*}\approx z(\Delta N+8) \,.
}
The IR fixed point approaches 1, with $w_\text{s}\to 0$,
\al{
v^\text{(IR)}_{0*}\approx 1-\frac{20z(1+w_\text{s})}{3(1-z\Delta N)}\, .
\label{IR v approx}
} The approximations \eqref{UV v approx}--\eqref{IR v approx} remain
valid for positive $\Delta N$ as long as $z\Delta N\ll 1$.

For the critical $z_\text{c}$ at which the fixed point disappears one
has \al{ v_\text{0c}&=\frac{1}{2}\left(1+z_\text{c} \Delta N\right)
  \,,& w_\text{s}\fn{v_\text{c}}&=\frac{2(1-z_\text{c}\Delta
    N)}{15\left(1+\frac{1}{6}(1-z_\text{c} \Delta N)\right)}\, ,& }
and therefore \al{ (1-z_\text{c}\Delta N)^2=\frac{80 z_\text{c}}{3}
  +\frac{32z_\text{c} (1-z_\text{c}\Delta
    N)}{9\left(1+\frac{1}{6}(1-z_\text{c} \Delta N)\right)}\,.
\label{relation of Delta N}
} 
For large $\Delta N$, this results in a value of $z_\text{c}\Delta N$
close to 1 such that the second term on the right-hand side of
\eqref{relation of Delta N} can be neglected. 
Therefore the critical boundary for $\tilde M_\text{p}^2$ linearly increases as a function of $\Delta N$,
\al{
  \tilde M_\text{p,c}^2\approx \frac{1}{64\pi^2}\left( \Delta N
    +\frac{4\sqrt{15}}{3}\sqrt{\Delta N}\right)\, .
\label{approximated Mpc}
}
The corresponding critical $v_\text{0c}$ approaches 1,
\al{
v_\text{0c}=1-\frac{2\sqrt{15}}{3\sqrt{\Delta N}}\,,
\label{approximated v0c}
} such that the graviton contribution is enhanced. The graviton
approximation becomes rather accurate for values of
$\tilde M_\text{p}^2$ and $v_0$ in the vicinity of \eqref{approximated
  Mpc} and \eqref{approximated v0c}.

\subsubsection{Critical exponent}
With the value of the UV fixed point, we obtain the critical exponent
of the cosmological constant or $v_0$ \al{ \theta_v&=-\frac{\p
    \beta_v}{\p v_0}\Bigg|_\text{at FP}=4-A\nn
  &=4-\frac{1}{12\pi^2{\tilde M}_{\text{p}*}^2}\left[ \frac{5}{(1-
      v_{0*})^2}+\frac{1}{4(1- v_{0*}/4)^2}\right]\nn
  &=\frac{4(1-2v_{0*})}{1-v_{0*}} \nn &\quad +\frac{1}{16\pi^2 \tilde
    M_{\text{p}*}^2(1-v_{0*})}\left( \Delta
    N+\frac{1}{(1-v_{0*}/4)^2}\right)\, .  } 
    Figure\,\ref{fig:critical
  exponent of cosmological constant} displays the dependence of
$\theta_v$ on the fixed-point value of the Planck mass for $N=4$ and
$\Delta N=0$. For the limit ${\tilde M}_{\text{p}*}\to \infty$ (weak
interaction), the critical exponent of the cosmological constant
asymptotically converges to 4, which is its canonical dimension.  On
the other hand, for ${\tilde M}_\text{p}\to {\tilde M}_\text{p,c}$ the
critical exponent approaches zero.

We display in Fig.\,\ref{fig:critical exponent of cosmological
  constant} various approximations. Besides the total contribution
(blue) and the graviton approximation (red), we show the total
contributions of the gravitational degrees of freedom without
contributions of other particles, i.e., $N=0$ and $\Delta N=0$. This
corresponds in \eqref{flow equation of U} to
$\tilde \pi_2+\tilde \pi_{0,g}+\tilde \eta$ (dashed green line). The
total gravitational contribution is again well approximated by the
graviton approximation. Finally, the green dashed line omits the
measure contribution $\tilde \eta$.

\begin{figure}
\includegraphics[width=\linewidth]{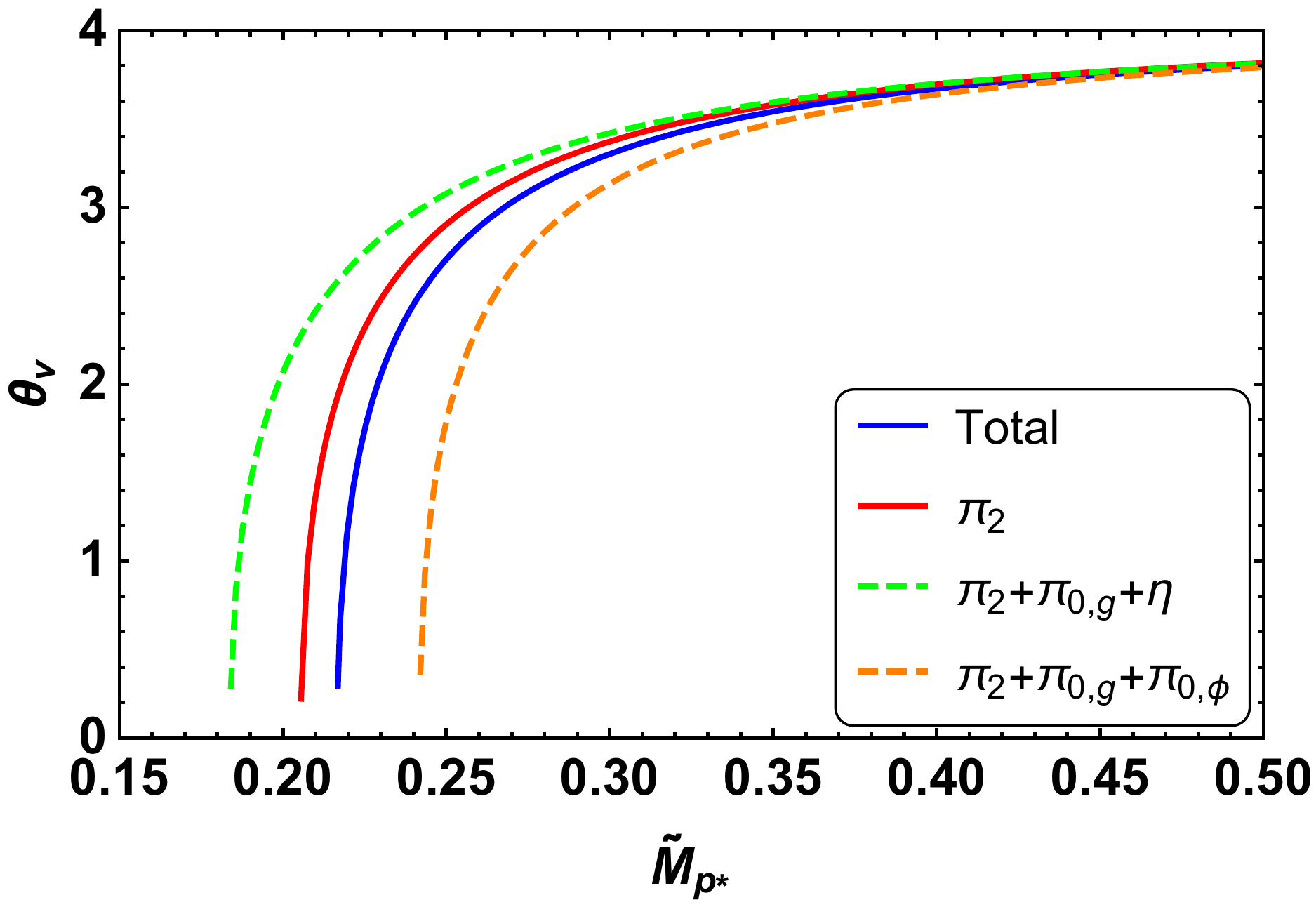}
\caption{The critical exponent of the cosmological constant $\theta_v$
  as a function of the fixed-point value of the Planck mass ($N=4$).
  The cosmological constant or, equivalently, $v_0$ is a relevant
  parameter. The individual contributions of various fluctuations are
  visualized by including in \eqref{flow equation of U} only the
  specified parts.  }
\label{fig:critical exponent of cosmological constant}
\end{figure}

\section{Predictions of asymptotic safety for the properties of the Higgs scalar}
\label{scalar interaction part}
As mentioned in the Introduction, a UV fixed point for quantum gravity
can predict those renormalizable couplings in the SM that correspond
to irrelevant couplings at the fixed point. We find that the quartic
coupling of the Higgs scalar is always irrelevant, and the scalar mass
term is irrelevant for a certain range of the fixed-point value of the
Planck mass.

\subsection{Effective low-energy theory}
So far, we have concentrated on the flow equations in the vicinity of
the UV fixed point of asymptotically safe quantum gravity. This
corresponds to a constant value of $\tilde M_\text{p}^2$ in the flow
equation \eqref{beta function of effective scalar potential}. Let us
now assume that the Planck mass corresponds to a relevant parameter in
quantum gravity. At short distances, it scales according to the
UV fixed-point behavior,
$M_\text{p}^2(k^2)=\tilde M_{\text{p}*}^2k^2$. For small $k$, it
deviates from this scaling behavior and takes a fixed value
$M^2$. This results in the qualitative behavior 
\al{
  M_\text{p}^2\fn{k}=
\begin{cases}
\tilde M_{\text{p}*}^2\, k^2 & \text{for $k>k_t$\,,}\\[5pt]
M^2 & \text{for $k<k_t$\,.}
\end{cases}\ }
 The transition scale is found as \al{
k_t^2=\frac{M^2}{\tilde M_{\text{p}*}^2}\,.  } A more complete
treatment smoothens the transition. Details of the threshold behavior
are not important for our purpose. One may use the simple form 
\al{
  M_\text{p}^2\fn{k}=M^2+\tilde M_{\text{p}*}^2k^2\,.  } 
  For $k$ below
the transition scale $k_t$, the dimensionless coupling
$\tilde M_\text{p}^2$ increases rapidly,
\al{ \tilde
  M_\text{p}^2=\frac{M_\text{p}^2}{k^2}=\tilde M_{\text{p}*}^2\left[
    \left( \frac{k_t}{k}\right)^2+1\right]\,.  } 
    As a result, the
gravitational contributions in \eqref{beta function of scalar mass}
and \eqref{beta function of quartic coupling}---not in \eqref{beta
  function of cosmological constant}---become rapidly tiny and can be
neglected. This leads to a simple picture. For $k<k_t$, the flow enters
the regime of an ``effective low-energy theory" for which the effect
of gravitational fluctuations can be neglected, except for the
cosmological constant $\tilde V$. This effective low-energy theory may
be the SM or a possible extension of it.

To rather good accuracy, the flow of dimensionless couplings as
$\tilde m_H^2$ and $\tilde \lambda_H$ can be divided into two
regimes. For the UV regime $k>k_t$, it follows the flow in the vicinity
of the UV fixed point. In contrast, for the IR regime $k<k_t$, the
flow is given by the low-energy effective theory. In this picture the
initial values of couplings for the IR flow, e.g., their values at
$k_t$, are determined by their final values of the UV flow. In case of
irrelevant couplings, the initial values for the IR flow are simply the
UV fixed-point values. They are therefore predicted. Following the IR
flow from $k_t$ in the vicinity of $M$ down to observable energy
scales leads then to predictions for observable quantities.

\subsection{Scalar mass term}
Next, we turn to the behavior of the scalar mass term in the vicinity
of the UV fixed point. Depending on the value of
$\tilde M_{\text{p}*}^2$, this can be a relevant or an irrelevant
parameter. This issue has important consequences for the gauge
hierarchy problem~\cite{Gildener:1976ai,Weinberg:1978ym}. Therefore,
we start from the discussion of this problem in the context of flow
equations.
\subsubsection{Scalar mass flow and gauge hierarchy problem}
Let us first discuss RG improved perturbation theory in the SM.
The RG equation of the scalar mass term at one-loop level is given by
\al{
\p_t\tilde  m_H^2=(-2+\gamma_m)\tilde m_H^2\,,
\label{RG equation of scalar mass simplied}
} where the first term on the right-hand side reflects the canonical
scaling and the second one is the anomalous dimension, which reads
\al{ \gamma_m = \frac{1}{16\pi^2}\left( 2\lambda_H
    +6y_t^2-\frac{9}{2}g^2-\frac{3}{2}g'{}^2\right)\,.
\label{perturbative anomalous dimension}
} More precisely, the coupling $\tilde m_H^2$ measures the distance
from the critical surface of the (almost) second-order vacuum
electroweak phase transition~\cite{Wetterich:1990an}. In
\eqref{perturbative anomalous dimension}, $\lambda$, $y_t$, $g$, and
$g'$ are the quartic coupling of the Higgs field, top-Yukawa coupling,
$\text{SU}\fn{2}$ gauge coupling, and $\text{U}\fn{1}$ gauge coupling,
respectively.  Using the values of the couplings at the Fermi scale,
the anomalous dimension $\gamma_m\approx 0.027$ is much smaller than
2.

For the marginal couplings in \eqref{perturbative anomalous dimension},
the scale dependence is logarithmic.  Neglecting their runnings the
solution of \eqref{RG equation of scalar mass simplied} reads \al{
  \tilde m_H^2={\tilde m}_0^2\left( \frac{k}{M}
  \right)^{-2+\gamma_m}\,,
\label{mass RG flow}
} where $\tilde m_0$ is the initial value of the scalar mass term at a
reference scale $M$.  For $M$ being of the order of the Planck scale,
one has to set a very tiny mass term $\tilde m_0^2\simeq 10^{-34} M^2$
at the Planck scale in order to obtain the Higgs mass
$\tilde m_H^2=m_H^2/\Lambda_\text{EW}^2\simeq 1$ at the electroweak
scale $k=\Lambda_\text{EW}\simeq {\mathcal
  O}\fn{10^{2}}\,\text{GeV}$. This is the gauge hierarchy problem.  It
is directly related to the role of $\tilde m_H^2=m_H^2/k^2$ being a
relevant coupling for the (approximate) fixed point of the SM, with
critical exponent $\theta_m=2-\gamma_m$.

A frequent discussion of the gauge hierarchy problem relies on the fact that the one-loop correction to the scalar mass involves a quadratic divergence as the UV cutoff is sent to infinity.
In perturbation theory, the observed Higgs mass is given by the cancellation between the squared bare mass and the quadratic divergence.
The quadratic divergence strongly depends on the cutoff scheme.
It is not present for dimensional regularization, while the momentum cutoff regularization and the Pauli-Villars type cutoff yield different values, depending on the precise implementation.

In terms of the RG, the quadratic divergence indicates the position of the phase boundary in the space of bare couplings. This boundary or ``critical surface" separates the symmetric and broken phases and corresponds to the massless (critical) situation~\cite{Wetterich:1983bi,Wetterich:1987az,Wetterich:2011aa,Wetterich:1990an,Wetterich:1991be,Aoki:2012xs,Wetterich:2016uxm}.
The position of the phase boundary depends on the precise definition and choice of the bare couplings and on the precise regularization. It changes under a coordinate change in ``theory space" if the latter is parametrized by the bare couplings. Different choices of the cutoff scheme also correspond to a coordinate transformation in theory space.
In quantum field theory, the precise choice of bare couplings is usually not of much interest.

On the other hand, the deviation from the phase boundary corresponds to a renormalized coupling. Its behavior is independent of the precise choice of microphysics as regularizations and the precise definition of bare couplings. This explains why the flow equation \eqref{RG equation of scalar mass} only involves renormalized couplings, while no trace of the quadratic divergence appears. The vanishing of the right-hand side reflects the basic property of a second-order phase transition. No trajectory can cross the phase boundary. Couplings on the critical surface stay on the critical surface.

\subsubsection{Quantum gravity effects}
We next add the effects of the gravitational quantum fluctuations in the range $k\gg M$.
The beta function of the scalar mass becomes
\al{
\beta_{m}&=-(2-\gamma_m-A)\tilde m_H^2\,.
}
Comparing with \eqref{RG equation of scalar mass simplied}, we see that $A$ corresponds to the gravitational contribution to the anomalous dimension.
Neglecting for simplicity the small value of $\gamma_m$ as compared to $A$, the critical exponent of the scalar mass parameter reads 
\al{
\theta_{m}&= -\frac{\p\beta_{m}}{\p \tilde m_H^2}\bigg|_\text{at FP}=2-A\nn
&= 2  - \frac{1}{48\pi^2 \tilde M_{\text{p}*}^2}\left[ \frac{20}{(1- v_{0*})^2}+\frac{1}{\left(1- v_{0*}/4\right)^2}\right]\,.
\label{critical exponent of scalar mass}
}
An important observation is that the sign of $A$ is positive. Gravitational fluctuations lower the value of the critical exponent $\theta_m$. As long as $A$ stays smaller than 2 ,the scalar mass term $\tilde m_H^2$ remains a relevant parameter, $\theta_m>0$. In this case, the distance from the vacuum electroweak phase transition, as measured by the value of the Fermi constant, cannot be predicted. It is simply a free parameter specifying the theory. A dramatic change occurs for $A>2$. In this event, the scalar mass term $\tilde m_H^2$ turns out to be an irrelevant coupling. The flow trajectory is always toward the phase---transition surface---an example of ``self-organized criticality." If asymptotically safe gravity is realized in a model leading to $A>2$, it predicts that $\tilde m^2\fn{k}$ vanishes for $k=M$. 
This explains the tiny value of the ratio $\tilde m^2\fn{k}/k^2=\tilde m_H^2$, evaluated at $k=M$, which is required by the observed Fermi scale, $\tilde m_H^2\fn{k=M}=10^{-34}$.
It produces an even stronger gauge hierarchy, namely $\tilde m_H^2\fn{k=M}=0$. 
If the vacuum electroweak phase transition would be an exact second-order phase transition, any model with $A>2$ would predict a vanishing Fermi scale.

The vacuum electroweak transition is not an exact second-order phase transition. This is due to the running gauge and Yukawa couplings that prevent the realization of exact scale symmetry in the effective low-energy theory below the Planck mass. The dominant effect is believed to be due to the running strong gauge coupling. Chiral symmetry breaking induces a quark-antiquark condensate $\langle \bar q q\rangle$. For the light quarks, this condensate sets a scale of the order $100$\,MeV, implying a lower bound on the Fermi scale of the same order of magnitude. The detailed effects of the scale violation in the top-quark--Higgs-scalar sector are not known quantitatively. If they are not substantially larger than the effects of the light quark condensate, asymptotically safe quantum gravity coupled to the SM predicts a $W$-boson mass around $100$\,MeV in case of $A>2$. This is not compatible with observation. 

For $A>2$, extensions would be needed, as discussed in the resurgence mechanism~\cite{Wetterich:2016uxm}. This may involve either new particles with masses near or below the Fermi scale or a more complicated UV fixed-point structure for quantum gravity. We recall in this context that a definition of the theory at the IR fixed point for $v_0$ (which becomes then the effective UV fixed point) implies $A>4$. The scalar mass term is always irrelevant for this setting.

\begin{figure}
\includegraphics[width=\linewidth]{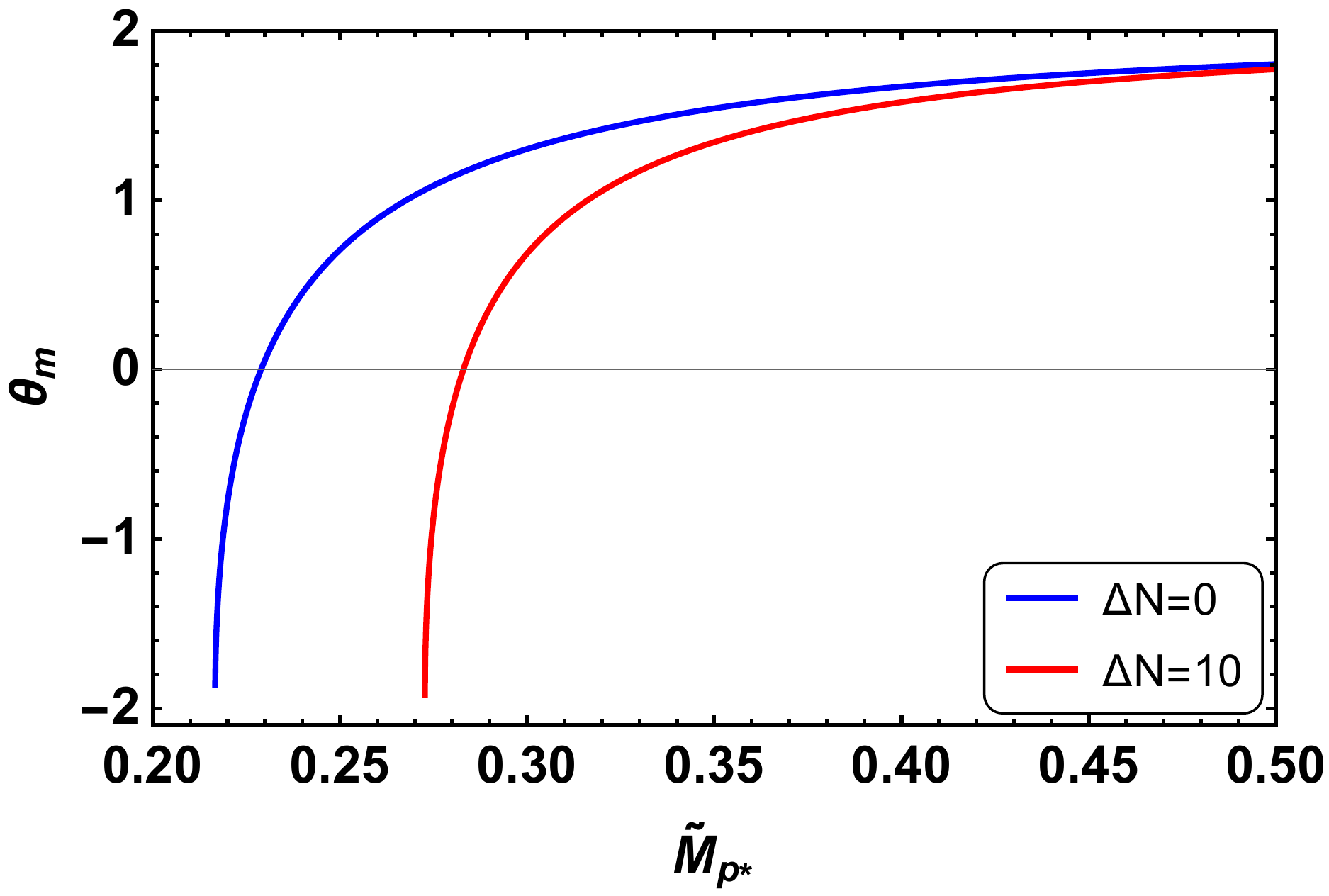}
\caption{The critical exponent of the scalar mass $\theta_m$ as a function of the fixed-point value of the Planck mass in the cases $\Delta N=0$ and $\Delta N=10$.
The mass term becomes irrelevant in the region where $\theta_m$ is negative (red-colored region in Fig.\,\ref{fig:scalar number and Planck mass}).
}
\label{fig:critical exponent of scalar mass}
\end{figure}

In Fig.\,\ref{fig:critical exponent of scalar mass}, we show the dependence of the critical exponent of the scalar mass term on the the fixed-point value of the Planck mass $\tilde M_\text{p}$ for $N=4$ and $\Delta N=0$. 
It falls below zero ($A>2$) if the strength of gravity exceeds a certain bound or $\tilde M_\text{p}$ becomes small enough. This behavior extends to other values of $\Delta N$, as shown in Fig.\,\ref{fig:scalar number and Planck mass} where $N$ stands for $4+\Delta N$. The red region in this figure occurs for $\theta_m<0$, $A>2$.

For large positive $\Delta N$, the physical scalar metric fluctuation can be neglected, yielding for the condition $A=2$ the relation
\al{
(1-v_0)^2=\frac{40z}{3}\, .
}
In this approximation, the fixed-point relation between $z$ and $1-v_0$ inferred from \eqref{simplified v equation} for $\p_t v_0=0$ reads
\al{
(1-v_0)+z\Delta N -1+\frac{20z}{3(1-v_0)}=0\,.
}
With
\al{
&(1-v_0)^2=(1-z\Delta N)^2-\frac{20z}{3} \\  
&\quad
-\frac{1}{2}(1-z\Delta N)\left[ 1-z\Delta N-\sqrt{(1-z\Delta N)^2 -\frac{80z}{3}}~\right] \,, 
\nonumber
}
and 
\al{
(1-z\Delta N)^2=30z\, ,
}
the line $A=2$ occurs for 
\al{
\tilde M_{\text{p}*}^2\approx \frac{1}{64\pi^2} \left( \Delta N+\sqrt{30\Delta N}\right)\,.
}
For large $\Delta N$, this comes very close to the critical value \eqref{approximated Mpc} for which the UV fixed point disappears. Only a small region with negative critical exponent $\theta_m$ remains.

\subsection{Quartic scalar coupling}
\begin{figure}
\includegraphics[width=\linewidth]{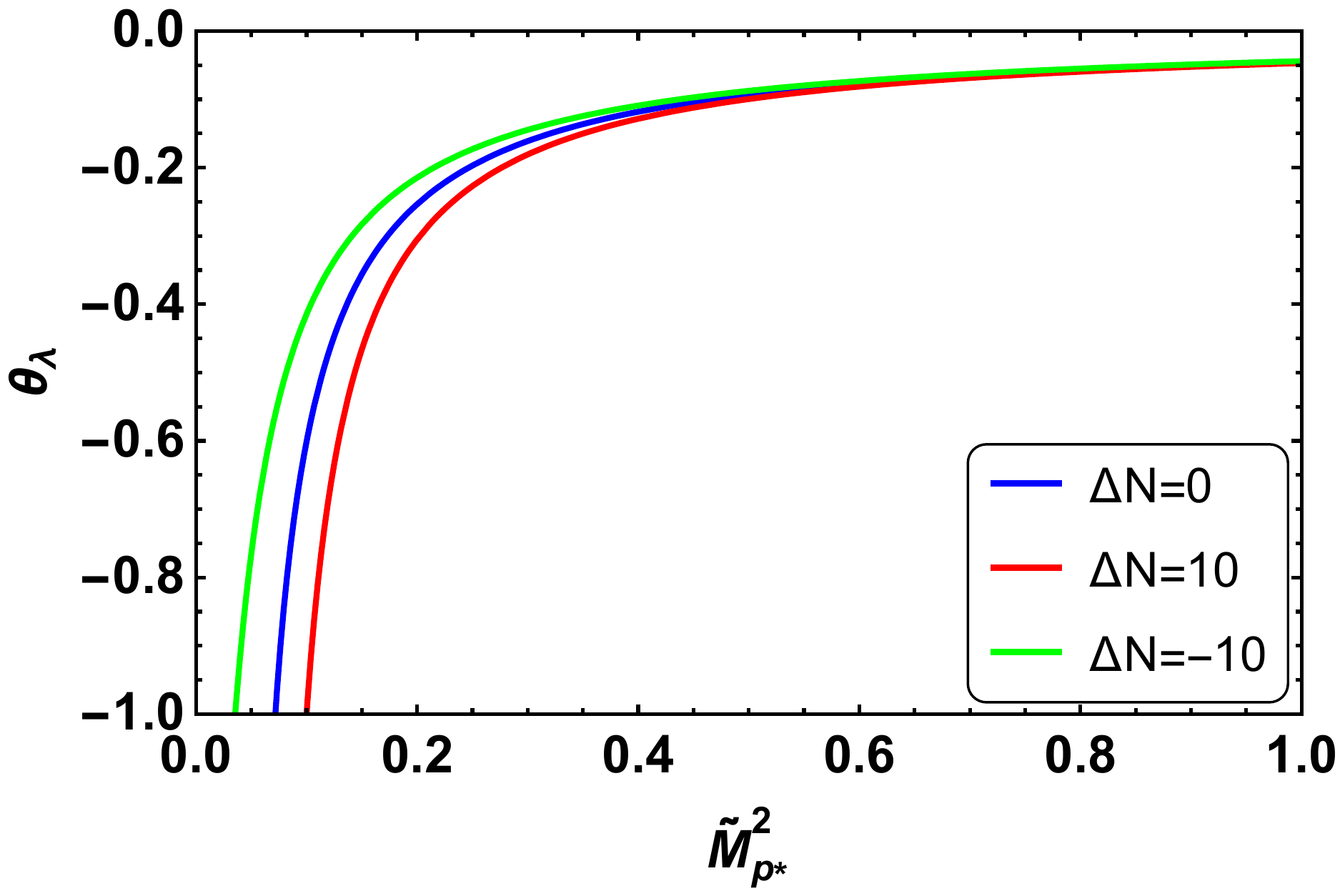}
\caption{The critical exponent of the quartic coupling $\theta_\lambda$ as a function of the fixed-point value of the squared Planck mass in the cases $\Delta N=0$, $\Delta N=10$, and $\Delta N=-10$.
}
\label{fig:critical exponent of quartic coupling}
\end{figure}

We finally investigate the quantum gravity effects on the quartic scalar coupling. 
The critical exponent of the quartic scalar coupling is given by the beta function in linear order in $\tilde \lambda_H$, $\beta_\lambda=-\theta_\lambda \tilde\lambda_H$,
\al{
\theta_\lambda&=-\frac{\p \beta_\lambda}{\p \tilde \lambda_H}\Bigg|_\text{at FP}=-A\nn
&=-\frac{1}{48\pi^2\tilde M_{\text{p}*}^2}\left[ \frac{20}{(1- v_{0*})^2}+\frac{1}{\left(1- v_{0*}/4\right)^2}\right]\,.
}
The dependence of the critical exponent on ${\tilde M}_\text{p}$ is shown in Fig.\,\ref{fig:critical exponent of quartic coupling}.

The irrelevance of the quartic coupling at the fixed point $\tilde \lambda_{H*}=0$ means that the coupling constant vanishes for $k$ above the Planck scale. In other words, the boundary condition of the RG equation of the quartic coupling is given by the fixed-point value $\tilde \lambda_{H*}=0$ at $k_t$ close to the Planck scale. With $\tilde \lambda_H\fn{k_t}=0$, the flow of $\tilde \lambda_H\fn{k}$ can be followed in the effective low-energy theory for $k<k_t$. Assuming that this IR flow is well approximated by the SM, the mass of the Higgs boson can be predicted as a function of the top-quark Yukawa coupling. The result of this prediction~\cite{Shaposhnikov:2009pv} was $m_H=126$\,GeV with only a few giga-electron-volts uncertainty. It agrees well with the observed value of the Higgs boson mass $m_H=125$\,GeV.

In the flow equation for $\tilde \lambda_H$, we have neglected effects of Yukawa couplings and gauge couplings. They contribute to the wave function renormalization, leading to a small anomalous dimension $\eta_\phi$. Yukawa and gauge couplings also shift the fixed point $\tilde \lambda_{H*}$ to a tiny nonzero value. 
For $\tilde \lambda_{H*}\neq 0$, the term proportional to $\tilde \lambda_H^2$ in $\beta_\lambda$ contributes to the anomalous dimension, as seen from the last term in \eqref{beta function of quartic coupling}.
This effectively enhances the anomalous dimension and makes the critical exponent $\theta_\lambda$ more negative. A nonzero fixed-point value $\tilde \lambda_{H*}$ could also be induced by nonminimal scalar interactions involving higher derivatives that may be generated by gravitational fluctuations.
In any case, even for nonzero $\tilde \lambda_H$ at the UV fixed point this value will be small. It therefore has only little impact on our results.

In summary, the negative sign of the critical exponent for the quartic scalar coupling seems to be a rather robust finding. For a given low-energy model the mass of the Higgs scalar is predictable in asymptotically safe quantum gravity.

\section{Robustness of results}\label{Robustness check}
In view of the far-reaching consequences of our findings for the predictivity of quantum gravity for SM parameters,
some tests of the robustness of these results seem appropriate.
Possible errors are connected with a possibly insufficient truncation of the exact flow equation.
Typical tests are the extension of the truncation and the sensitivity to the choice of the infrared cutoff function.
Gauge dependence is a minor issue in our approach since we are bound to employ a physical gauge fixing that only acts on the gauge degrees of freedom in the metric.
Within this class of physical gauge fixings, the dependence on the precise gauge fixing is small~\cite{Wetterich:2016ewc,Wetterich:2017aoy}.
At the present stage, the main uncertainty concerns the fixed-point value of the dimensionless Planck mass. 
We will see that truncation errors can typically be compensated by a change of this value.
The error analysis will become more meaningful at a later stage when the fixed-point value for the Planck mass is also calculated. 
Nevertheless, our two main statements seem rather robust: 
(i) The quartic scalar coupling is an irrelevant parameter at the UV fixed point. 
(ii) There exists a range of fixed-point values for the Planck mass for which the scalar mass term is also an irrelevant parameter.

\subsection{Extension of truncation}
So far, we have analyzed the Einstein-Hilbert truncation for the gravity sector.
Extensions of the truncation are possible in various directions.
They are, however, restricted by diffeomorphism symmetry of the effective action.
One possibility is the inclusion of higher-order curvature invariants.
Only the quadratic invariants influence the propagator of the metric fluctuations in flat space, which is the only quantity involved for the computation of the flow of the effective scalar potential.
From the quadratic curvature invariants, it is only the squared Weyl tensor that enters in the transverse traceless mode of the metric  propagator, which constitutes the dominant graviton contribution.
Denoting by $D$ the coefficient of the squared Weyl tensor, the inverse graviton propagator becomes at the fixed point
\al{
G_g^{-1}=\frac{\tilde M_{\text{p}*}^2}{4}\left( k^2 q^2-k^4 v_{0*}\right) +\frac{D_*}{2} q^4 \,.
}
The flow is dominated by momenta $q^2\approx k^2$, such that $D$ becomes important only for
\al{
D_* \gtrsim \frac{\tilde M_{\text{p}*}^2}{2}(1-v_{0*})\,.
}
A positive $D_*$ lowers the anomalous dimension $A$, while a negative $D_*$ enhances it.
The sign of $A$ is not changed as long as gravity is stable for positive $G_g^{-1}$ in the Euclidean domain.
The inclusion of $D$ modifies the factor $(1-v_0)^{-2}$ in the first term of \eqref{anomalous dimension A} to $[(1-v_0)+2 D/ \tilde M_\text{p}^2]^{-2}$.
For large $D$ this replaces $\tilde M_\text{p}^{-2}(1-v_0)^{-2}$ by $\tilde M_\text{p}^2/(4D^2)$.  
The crucial property $A>0$ also holds within the perturbatively renormalizable quartic gravity~\cite{Stelle:1976gc,Salvio:2014soa}.

\begin{figure*}
\includegraphics[width=8.5cm]{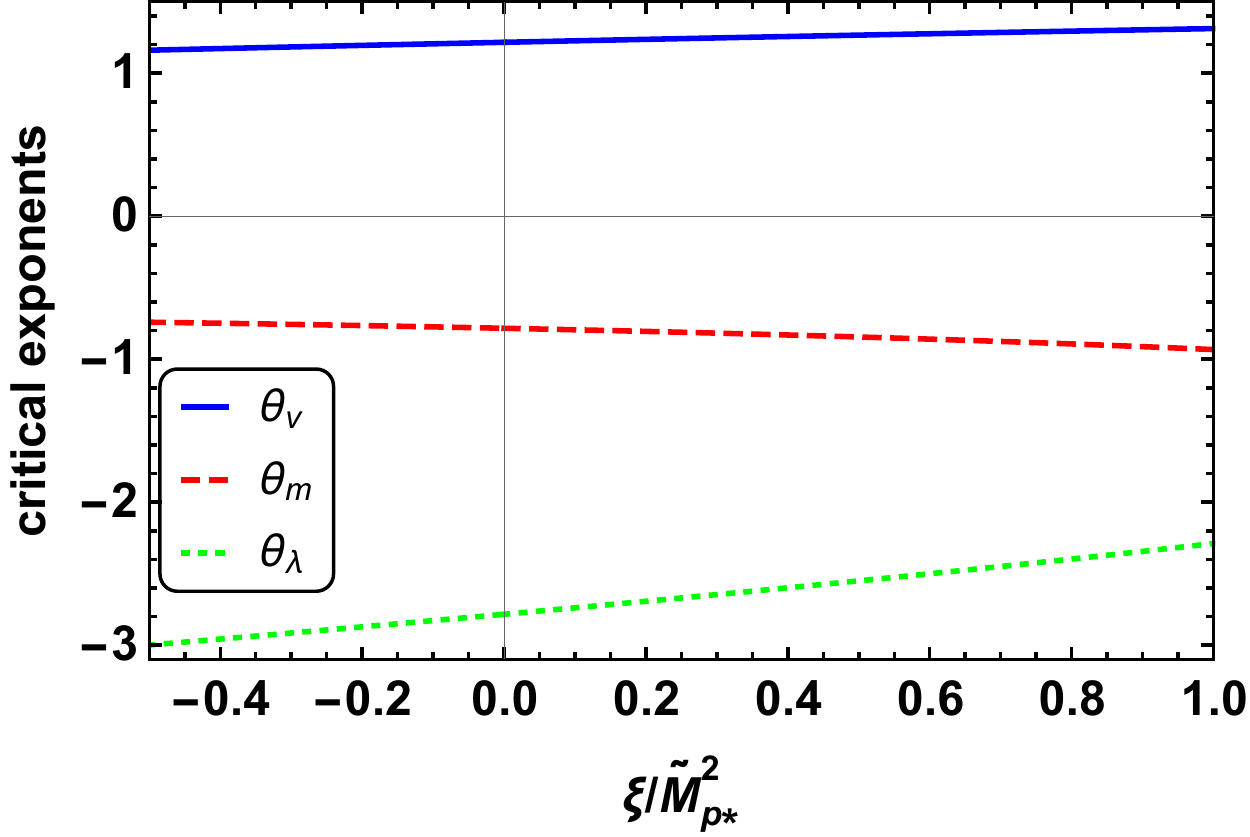}
\includegraphics[width=9cm]{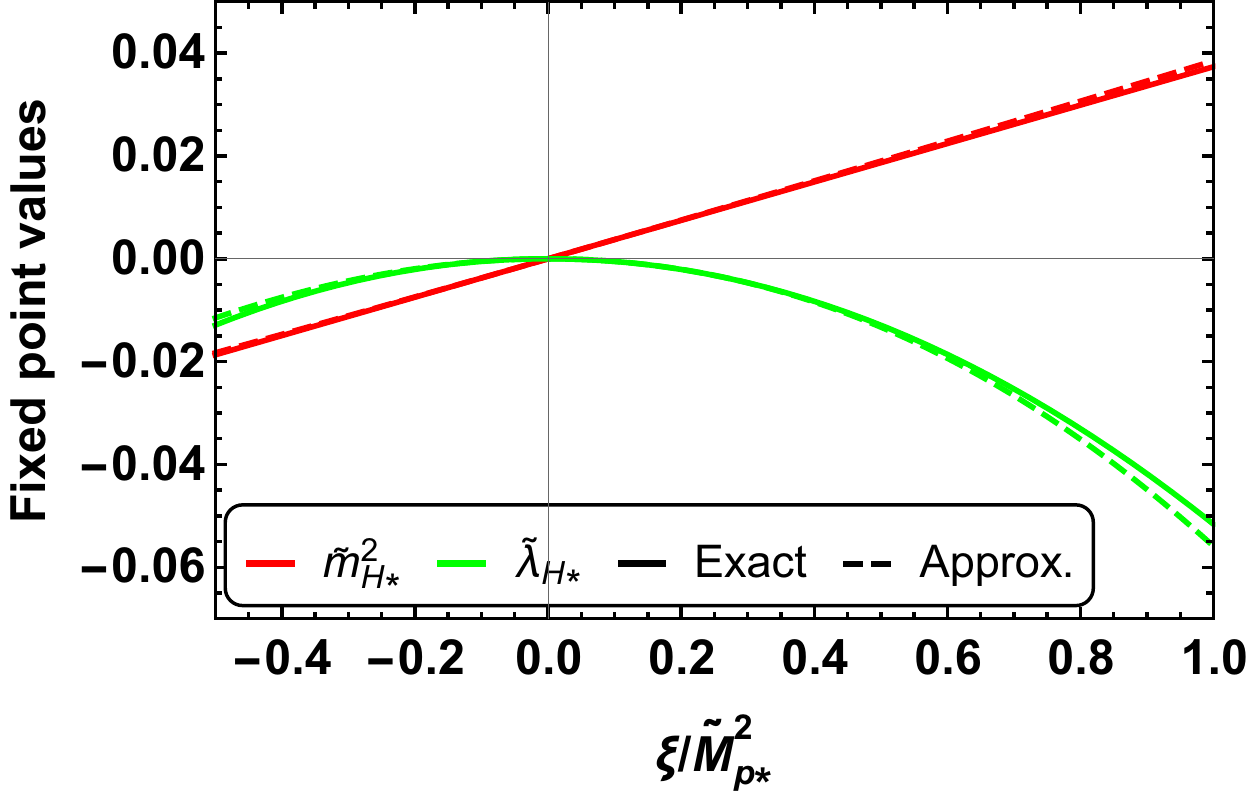}
\caption{
Critical exponents (left) and fixed-point values (right) of the scalar mass term and quartic coupling as functions of the nonminimal coupling $\xi/\tilde M_{\text{p}*}^2$. For the dimensionless Planck mass, we take the value $\tilde M_\text{p*}=0.22$.
In the right-hand panel, we compare the fixed-point values for the numerical solutions including all terms discussed here (solid line) and the approximated ones given in \eqref{Fixed point for lambda} (dashed line).
Importantly the critical exponents, displayed in the left-hand panel, are almost independent of the nonminimal coupling $\xi/\tilde M_{\text{p}*}^2$.
}
\label{fig:dependence of anomalous dimension on nonminimal coupling} 
\end{figure*}

For a second point, we investigate a possible dependence of the effective Planck mass on the scalar field, extending the truncation by a nonminimal interaction between the scalar field and the curvature,
\al{
\Gamma_k=-\frac{1}{2}\int_x \sqrt{g}\left(M_\text{p}^2 +\xi \rho \right) R \,.
} 
We consider here a constant $\xi$, i.e., assumed to be at its fixed-point value.
In the following we derive the shift in the fixed-point values of $m_H$ and $\lambda_H$, 
the changes in the stability matrix as well as new contributions due to the mixing in the scalar sector.
For a main result, we find that the dependence of the critical exponents on the nonminimal coupling is small 
as displayed in Fig.\,\ref{fig:dependence of anomalous dimension on nonminimal coupling}.

The flow equation \eqref{pi0 contribution} for the effective potential remains valid if we replace $\tilde M_\text{p}^2\to \tilde M_\text{p}^2+\xi \tilde \rho$, $v=2\tilde U/(\tilde M_\text{p}^2+\xi \tilde \rho)$.
The additional dependence of $v$ on $\tilde \rho$,
\al{
\frac{\p v}{\p \tilde \rho}=\frac{2}{\tilde M_\text{p}^2+\xi \tilde \rho}\left( \tilde U' -\frac{ \xi v}{2}\right) \,,
}
modifies the flow equation for the mass term $\tilde m_H^2=\p \tilde U/\p \tilde \rho|_{\tilde \rho =0}$ and the quartic coupling 
$\lambda_H =\p^2\tilde U/\p \tilde \rho^2|_{\tilde \rho =0}$ \cite{Wetterich:2019qzx}.
On the other hand, the flow equation for $v_0=2\tilde V/\tilde M_\text{p}^2$ depends on $\xi$ only indirectly through the dependence of the scalar contribution $\pi_0$ on $\tilde m_H^2$.

We first neglect the mixing between the scalar modes, which will be added below.
The flow equation \eqref{RG equation of scalar mass} for $\tilde m_H^2$ gets extended to 
\al{
\p_t \tilde m_H^2=(-2+A)\tilde m_H^2-\frac{ \xi A v_0}{2} -\frac{3\tilde \lambda_H}{16\pi^2(1+\tilde m_H^2)^2}\,.
\label{extended flow equation of m}
}  
Similarly, the flow equation for $\lambda_H$ \eqref{RG equation of quartic coupling} becomes
\al{
\p_t \tilde \lambda_H&=A\tilde \lambda_H +\frac{2}{\tilde M_\text{p}^2}\frac{\p A}{\p v}\left( \tilde m_H^2 -\frac{\xi v_0}{2} \right)^2
\nn &\quad 
-\frac{2\xi A}{\tilde M_\text{p}^2}\left( \tilde m_H^2 -\frac{\xi v_0}{2} \right)+\frac{3\tilde \lambda_H^2}{4\pi^2(1+\tilde m_H^2)^3}\,,
\label{extended flow equation of lambda}
}
where all quantities are evaluated at $\tilde \rho=0$ and we take $N=4$ for the number of scalars in \eqref{beta function of quartic coupling}.

The fixed-point value of $\tilde m_H^2$ is now nonvanishing (assuming $A\neq 2$),
\al{
\tilde m_{H*}^2=\frac{1}{A-2}\left( \frac{\xi Av_{0*}}{2}+\frac{3\tilde \lambda_{H*}}{16\pi^2(1+\tilde m_{H*}^2)^2}\right)\,.
\label{m fixed point}
}
For positive $\xi$ and positive or small $\tilde \lambda_{H*}$, the origin at $\tilde \rho=0$ is a local minimum only for $A>2$.
Similarly, the fixed-point value for $\tilde \lambda_H$ is also nonzero.
A negative value of $\tilde \lambda_H$ is not an indication of instability.
The Taylor expansion is not valid for larger values of $\tilde \rho$ for which $U$ would get negative in a quartic approximation.
If the last term $\sim \tilde \lambda_H^2$ in \eqref{extended flow equation of lambda} can be neglected and if one can approximate $\tilde m_{H*}^2\ll 1$ in \eqref{m fixed point}, one has
\al{
\tilde \lambda_{H*}=\frac{C}{\tilde A} \,,
}
with 
\al{
C=\frac{8\tilde m_{H*}^4}{A v_{0*} \tilde M_\text{p*}^2}\left( A-2 -\frac{\p \ln  A}{\p \ln v}\right) \,,
}
and
\al{
\tilde A=A+\frac{3\tilde m_{H*}^2}{4\pi^2Av_{0*} \tilde M_\text{p*}^2}\left(4 -A +2\frac{\p \ln A}{\p\ln v}\right)\,.
}
This approximation breaks down as $\tilde A$ comes close to zero.
For the full fixed-point solution of Eqs.\,\eqref{extended flow equation of m} and \eqref{extended flow equation of lambda}, both $\tilde m_{H*}^2$ and $\tilde \lambda_{H*}$ remain finite.
For small $\xi$, one has $\tilde m_{H*}^2\sim \xi$, $\tilde \lambda_{H*}\sim \xi^2$, such that in lowest order in $\xi$ one obtains
\al{
\tilde m_{H*}^2&=\frac{\xi Av_{0*}}{2(A-2)} \,,\nn
\tilde \lambda_{H*}&=\frac{2\xi^2 v_{0*}}{\tilde M_{\text{p}*}^2}\frac{A-2-\frac{\p \ln  A}{\p \ln  v}}{(A-2)^2} \,.
\label{Fixed point for lambda}
}
For the graviton approximation one may use the simple relation 
\al{
\frac{\p\ln A}{\p \ln v}=\frac{2v_0}{1-v_0} \,.
}

The stability matrix $T$ receives additional off-diagonal entries for $\tilde m_{H*}^2\neq 0$, $\tilde \lambda_{H*}\neq 0$,
\al{
T=\pmat{
4-A &  \frac{1}{4\pi^2 \tilde M_\text{p*}^2(1+\tilde m_{H*}^2)^2} & 0\\[10pt]
B & 2-A -\frac{3\tilde \lambda_{H*}}{8\pi^2(1+\tilde m_{H*}^2)^3}& \frac{3}{16\pi^2(1+\tilde m_{H*}^2)^2}\\[10pt]
E & F & -A -\frac{3\tilde \lambda_{H*}}{2\pi^2(1+\tilde m_{H*}^2)^3}
},\label{extended stability matrix}
}
where
\al{
B&=\frac{\xi A}{2}\left( 1+\frac{\p \ln A}{\p\ln v}\right) -\frac{\p A}{\p v}\tilde m_{H*}^2\nn
&=\frac{\xi A}{2}\left( 1-\frac{2}{A-2}\frac{\p \ln A}{\p\ln v}\right)\,,
\label{B factor}
}
and
\al{
F&=\frac{1}{\tilde M_{\text{p}*}^2}\left[2 \xi A-4 \frac{\p A}{\p v}\left( \tilde m_{H*}^2 -\frac{\xi v_{0*}}{2}\right)\right]
+\frac{9\tilde \lambda_{H*}^2}{4\pi^2(1+\tilde m_{H*}^2)^4}
\nn
&= \frac{2\xi A}{\tilde M_{\text{p}*}^2}\left(1-\frac{2}{A-2}\frac{\p \ln\,A}{\p \ln\,v} \right)\,.
\label{F factor}
}
Here the second lines in \eqref{B factor} and \eqref{F factor} use the approximation \eqref{Fixed point for lambda}.
For $E$, one obtains
\al{
E&=\frac{2}{\tilde M_{\text{p}*}^2}\Bigg[ 2 \xi \frac{\p A}{\p v}\left( \tilde m_{H*}^2-\frac{\xi v_{0*}}{2}\right) -\frac{\xi^2 A}{2} \nn
&\quad -\frac{\p^2 A}{\p v^2} \left( m_{H*}^2-\frac{\xi v_{0*}}{2} \right)^2 \Bigg] 
-\frac{\p A}{\p v}\lambda_{H*}\,,
}
which is of the order $\xi^2$.
The corrections from $\xi$ are small as long as
\al{
\frac{|B|}{4\pi^2 \tilde M_{\text{p}*}^2}&< |(4-A)(2-A)| \,,\nn
\frac{3|F|}{16\pi^2}&< |A(2-A)| \,.
}
For an order of magnitude estimate, this holds for
\al{
|\xi| < |A-2|\pi^2 \tilde M_\text{p*}^2\, .
}

Since for $\xi\neq 0$ the scaling solutions for $\tilde m^2_{H*}$ and $\tilde \lambda_{H*}$ occur for nonzero values, there is also a contribution to the flow equation from the mixing in the scalar sector.
This only concerns a subleading term.
For small $\tilde \rho\, \tilde U'{}^2/(\tilde M_\text{p}^2+\xi \tilde \rho)$, the mixing contributes an additional term to $\tilde \pi_0$,
\al{
\Delta (\p_t \tilde U)=\frac{\Delta\tilde \pi_0}{k^4}=-\tilde \rho \tilde U'{}^2H\,,\nonumber
}
with
\al{
H=\frac{1+\tilde U'+2\tilde \rho \tilde U'' +\frac{3}{4}(1-\frac{v}{4})}{8\pi^2 (\tilde M_\text{p}^2+\xi \tilde \rho)(1-\frac{v}{4})^2(1+\tilde U'+2\tilde \rho \tilde U'')^2}\,,
}
where we take $\eta_g=\eta_\phi=0$.
This does not contribute to the flow of $V$ or $v_0$, but the contribution to the flow of $\tilde m_{H}^2$ and $\tilde \lambda_H$  vanishes only for $\tilde m_{H}^2=0$, $\tilde\lambda_H=0$. 
One finds, with $H_0=H\fn{\tilde\rho=0}$,
\al{
\Delta (\p_t \tilde m_H^2)=-H_0 \tilde m_{H}^4 \,,
}
and 
\al{
\Delta (\p_t \tilde \lambda_H)= -2H_0 \tilde m_H^2 \tilde \lambda_H -\tilde m_H^4\frac{\p H}{\p \tilde \rho}\bigg|_{\tilde \rho=0}\,.
}
For small $\xi$, this shifts the fixed-point value $\tilde m_{H*}^2$ by a small amount approximately $\xi^2$, while the shift in $\tilde\lambda_{H*}$ is approximately $\xi^3$.
In leading order in $\xi$ these shifts can be neglected.

The dominant contribution of the mixing effect to the stability matrix $T$ is a shift in the diagonal terms for $\delta \tilde m_H^2$ and $\delta\tilde \lambda_H$ (not for $\delta v$),
\al{
A\to A-2H_0 \tilde m_{H*}^2\,.
\label{extended A}
}
As discussed before, it vanishes for $\xi\to 0$, $\tilde m_{H*}^2\to 0$.
With
\al{
H_0=\frac{1+\tilde m_{H*}^2+\frac{3}{4}\left( 1-\frac{v_{0*}}{4}\right)}{8\pi^2\tilde M_{\text{p}*}^2\left(1-\frac{v_{0*}}{4}\right)^2(1+\tilde m_{H*}^2)^2}\,,
}
we may neglect in leading order $\tilde m_{H*}^2$ in $H_0$ and employ Eq.\,\eqref{Fixed point for lambda} for $\tilde m_{H*}^2$ in \eqref{extended A}.
We observe that $H_0$ is positive such that $A$ is enhanced for negative $\tilde m_{H*}^2$.
For $A<2$, one has $\tilde m_{H*}^2<0$ such that the critical exponent for the scalar mass term moves closer to zero by the mixing effect.

The inclusion of the nonminimal coupling $\xi$ further modifies the off-diagonal parts in the inverse propagator of the spin-0 metric fluctuation in \eqref{physical modes matrix}, such that 
\al{
\frac{1}{2}{U' \phi }\to \frac{1}{2}(-\xi q^2+U')\phi \,.
}
This effect induces a shift in the anomalous dimension, adding to \eqref{extended A} a further piece
\al{
A\to A-2H_0 \tilde m_{H*}^2+\Delta A_i \,,
\label{shift of A}
}
with
\al{
\Delta A_v&=0\,,\qquad
\Delta A_m=\frac{3\xi(\xi+2)}{32\pi^2\tilde M_\text{p}^2(1-v_0)^2}\,,\nn
\Delta A_\lambda&=\frac{3\xi(3\xi+2)}{16\pi^2\tilde M_\text{p}^2(1-v_0)^2}\,.
\label{deltaA}
}
Even if the fixed-point values of $\tilde m_{H*}^2$ and $\tilde \lambda_{H*}$ are zero, this modification differs from zero as long as $\xi\neq 0$.

A characteristic quantity for the field dependence of the effective Planck mass is $\xi/\tilde M_\text{p}^2$.
For $\xi/\tilde M_\text{p}^2\ll 1$, the field dependence is weak, while for $\xi/\tilde M_\text{p}^2\gsim 1$, the Planck mass varies very rapidly with the scalar field.
Figure\,\ref{fig:dependence of anomalous dimension on nonminimal coupling} displays the dependence of the critical exponents on $\xi/\tilde M_{\text{p}*}^2$ by evaluating the eigenvalues of the stability matrix \eqref{extended stability matrix} including the shift \eqref{shift of A} in the diagonal elements.
We use the value of the dimensionless Planck mass $\tilde M_{\text{p}*}=0.22$ at which we have the anomalous dimension $A\approx 2.78$ for vanishing $\xi$.
The dominant effect arises from the nonzero values of $\tilde m_{H*}^2$ and $\tilde \lambda_{H*}$.
In a realistic setting, also gauge and Yukawa couplings influence these fixed-point values.
If the difference of $\tilde m_{H*}$ and $\tilde \lambda_{H*}$ from zero can be neglected, the effect of $\xi$ is much smaller, given by \eqref{deltaA}.
For a main result, we find that the dependence of the critical exponents on $\xi$ is small, supporting our conclusions from the previous sections.

\subsection{Regulator dependence}
The functional flow equation is exact for an arbitrary choice of the regulator function $\mathcal R_k$. 
Provided that $\mathcal R_k$ obeys the requirements for an efficient IR cutoff, with $\p_t \mathcal R_k$ decaying fast for high momenta, the results for observable quantities should not depend on the choice of $\mathcal R_k$.
In any practical calculation, they do, however, and this is due to the choice of a truncation.
This observation can be used for a test of validity of a given truncation.
The dependence on $\mathcal R_k$ should disappear for a ``perfect truncation,'' and any remaining dependence can be taken as some form of measure for the error induced by the truncation.

For a general cutoff function, the graviton contribution to the flow of the potential is given by the threshold function $\ell_0^4$,
\al{
\tilde \pi_2=\frac{5k^4}{16\pi^2}\ell_0^4\fn{-v_0},
}
with $\ell_0^4\fn{\tilde w}$ defined in terms of the propagator $G\fn{q^2}$ as
\al{
\ell_0^4\fn{\tilde w}&=\frac{8\pi^2}{k^4}\int_q\p_t\mathcal  R_k\fn{q^2} \,G\fn{q^2}\nn
&=\frac{1}{2}\int^\infty_0 \df x\,x(p\fn{x}+\tilde w)^{-1}f\fn{x}r\fn{x},
}
with $x=q^2/k^2$ and
\al{
p\fn{x} &=x+r\fn{x}\,,
\quad
r_k\fn{x}=\frac{4\mathcal R_k}{M_\text{p}^2k^2} \,,
\quad
\eta_M=\frac{\p_t M_\text{p}^2}{M_\text{p}^2} \,, &
\nn
f_k\fn{x}&=\frac{\p_t\mathcal R_k}{\mathcal R_k}=2+\eta_M-2\frac{\p \ln\,r}{\p\ln\, x} \,.
}
For the Litim cutoff, one takes
\al{
{\mathcal R}_k=\frac{M_\text{p}^2}{4}(k^2-q^2)\theta\fn{k^2-q^2},
}
with
\al{
f_k\fn{x}r_k\fn{x}=2+\eta_M(1-x)
}
and
\al{
p\fn{x}=\begin{cases}
1 & \text{for}~x<1\,,\\
x & \text{for}~x>1\,.
\end{cases}
}
This results in the threshold function
\al{
\ell_0^4 (\tilde w) =\left( \frac{1}{2}+\frac{\eta_M}{12}\right)\left(1 +\tilde w\right)^{-1}\,.
}
For $\eta_M=2$, corresponding to $\eta_g=0$, we recover \eqref{pi0 contribution}.

For a general cutoff function, one replaces in the flow equation for $\tilde U$
\al{
(1-v_0)^{-1}\to \frac{3}{2}\ell_0^4\fn{-v_0} \,,
}
and a similar replacement is performed for the scalar contributions.
For the anomalous dimension $A$, one has to replace 
\al{
(1-v_0)^{-2}\to \frac{3}{2}\ell_1^4\fn{-v_0} \,,
}
with
\al{
\ell_1^4\fn{\tilde w}=-\frac{\p \ell_0^4\fn{\tilde w}}{\p \tilde w} \,.
}
Since $A$ is dominated by the graviton contribution a good estimate of the effect of a general cutoff function is the multiplication of Eq.\,\eqref{anomalous dimension A} by a factor $f$,
\al{
f=\frac{3}{2}\ell_1^4\fn{-v_0}(1-v_0)^2 \,.
}
For the Litim cutoff, $f=1$.
Correspondingly, the value of $\tilde M_{\text{p}*}^2$ needed to realize a given $A$ has to be multiplied by $f$.
This feature is not surprising.
A dominant effect of a change in the IR-cutoff function can be viewed as an effective rescaling of $k$, which may be absorbed by a multiplicative redefinition of $k$.
The dimensionless ratio $\tilde M_\text{p}^2=M_\text{p}^2/k^2$ is directly affected by such a rescaling.
This property demonstrates that our general results hold independently of the precise choice of the cutoff function.

\begin{figure}
\includegraphics[width=\linewidth]{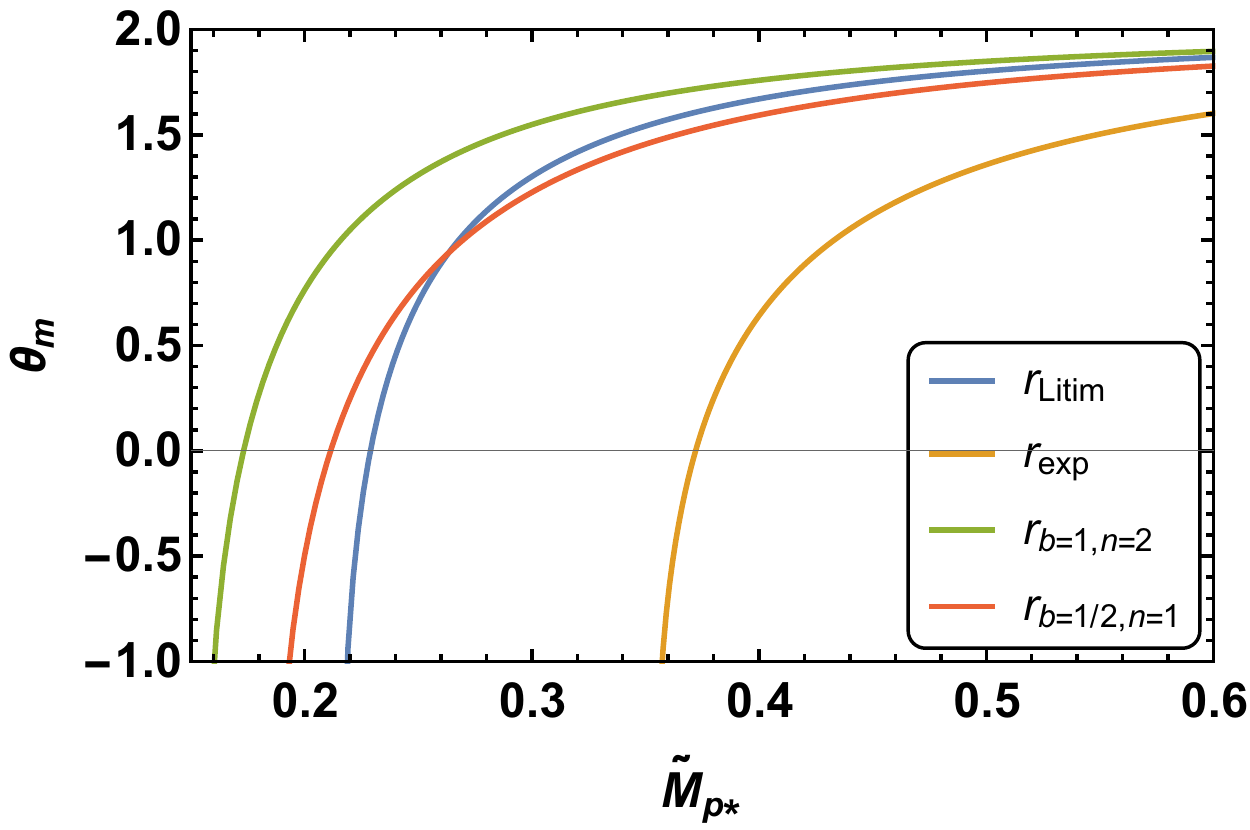}
\caption{
Critical exponent of the scalar mass term as a function of the fixed-point value of the Planck mass for different regulators.
The Litim regulator is depicted in blue, the exponential regulator is in yellow, $r_\text{int}(x,\,b=1,\, n=2)$ is in green, and $r_\text{int}(x,\,b=\frac12,\, n=1)$ is in red.
The main effect is the expected multiplicative factor in the value of $\tilde M_{\text{p}*}$.
}
\label{fig:RegDependence} 
\end{figure}

Let us investigate the regulator dependence numerically. 
To this end, we use an exponential interpolating cutoff function for the explicit regulator comparison
\al{
 r_\text{int} (x,\, b,\, n) = \frac{(1 - b\, x ) x^{n-1}}{\exp(x^n)-1} \,.
}
This cutoff function has the limits $r_\text{int}(x,\,b=1,\, n\to \infty) = r_\text{Litim} (x) = (x^{-1}-1)\theta(1-x)$ 
as well as $r_\text{int}(x,\,b=0,\, n=1) = r_\text{exp} (x) = (\exp(x)-1)^{-1}$.
We further use the regulator at the values $r_\text{int}(x,\,b=1,\, n=2)$ as well as $r_\text{int}(x,\,b=\frac12,\, n=1)$.
The results are displayed in Fig.~\ref{fig:RegDependence}.
One can see that the fact that the critical exponent of the scalar mass $\theta_m=2-A$ becomes negative for a certain value of the dimensionless Planck mass is not changed.
The dominant effect is indeed a simple rescaling of $\tilde M_{\text{p}*}^2$.
For a given model and a given truncation, $\tilde M_{\text{p}*}^2$ will depend sensitively on the chosen cutoff function.
For a valid truncation, this cutoff dependence should drop out in the final value or $\theta_m$.
A computation of the flow equation for $\tilde M_\text{p}^2$ will be needed for this check.

\section{Discussion}
\label{discussion section}
We have computed quantum gravity predictions for the mass and couplings of the Higgs scalar within the asymptotic safety scenario. We consider the SM of particle physics coupled to gravity, with possible extensions of the particle content. The value of the dimensionless flowing Planck mass at the fixed point, $\tilde M_\text{p}^2=M_\text{p}^2\fn{k}/k^2$, is influenced by the particle content of the model. We treat it here as an unknown parameter, to be determined for any given model.

Our main findings are the following: (i) The quartic self-coupling $\tilde \lambda_H$ of the Higgs scalar is an irrelevant coupling. Its value at the Planck scale is predicted to be very close to zero. For a given low-energy model below the Planck scale, where gravitational contributions decouple, this initial value at the Planck scale is mapped by the renormalization flow to the value at the Fermi scale. The ratio between Higgs-boson mass and top-quark mass is therefore predicted. This prediction works well if the low-energy theory is the SM. The consistency of other low-energy models has to be tested. (ii) The dimensionless mass term $\tilde m_H^2$ for the Higgs scalar can be a relevant or an irrelevant coupling, depending on the fixed-point value $\tilde M_\text{p}^2$ and on the degrees of freedom, as shown in Fig.\,\ref{fig:scalar number and Planck mass}. If $\tilde m_H^2$ is relevant, the value of the Fermi scale, or more precisely the ratio between Fermi scale and Planck scale $M_W/M_\text{p}$, cannot be predicted.
For a relevant $\tilde m_H^2$, the gauge hierarchy is a free parameter. In contrast, if $\tilde m_H^2$ is irrelevant, the model predicts that nature is located on the critical surface of the vacuum electroweak phase transition, with only a small deviation induced by running gauge and Yukawa couplings. Depending on the model, this may overpredict the gauge hierarchy to be $10^{-40}$ instead of $10^{-34}$. (iii) For a given particle content, as expressed by $N$, we find a lower bound on the fixed-point value of $\tilde M_\text{p}^2$ if $N>-4$.
  
These results extend to other quantum field theories with scalar fields coupled to gravity, such as extensions of the SM or grand unified theories. The gravitational contribution to the critical exponents is universal. Quartic scalar couplings are irrelevant parameters at the UV fixed point.

Our findings are of high relevance for the interplay between quantum gravity and particle physics. The validity of approximations and the robustness of results should therefore be critically questioned.  First of all, quantum gravity contributions to the effective scalar potential can be performed in flat space, allowing for the full use of Euclidean SO(4) symmetry or Lorentz symmetry for Minkowski space. Second, our split of gravitational fluctuations into physical modes and gauge modes, together with a physical gauge fixing acting only on the gauge modes, makes the contributions of different modes very transparent. We find that the dominant contributions in the gravitational sector come from the fluctuations of the graviton or traceless transverse tensor fluctuations. 

The quantity needed for a reliable computation of the dominant graviton contribution is the exact propagator of the graviton. This is not directly available, and at this point an approximation is made. In the Einstein-Hilbert truncation employed in the present paper, the inverse graviton propagator is given by 
\al{
G^{-1}\fn{q^2}=\frac{M_\text{p}^2 q^2}{4}-\frac{U}{2}\,,
}
and the question arises if this is a reasonable approximation. The inverse graviton propagator at zero momentum is given by the effective potential 
\al{
G^{-1}\fn{q^2=0}=-\frac{U}{2}\,.
}
For a diffeomorphism invariant formulation, this is an exact relation enforced by diffeomorphism symmetry~\cite{Wetterich:2018poo}. The right-hand side of the flow equation involves a momentum integral that is dominated by momenta with $q^2\approx k^2$. We may therefore {\it define} the parameter $M_\text{p}^2\fn{k}$ by the graviton propagator at $q^2=k^2$, more precisely by 
\al{
M_\text{p}^2\fn{k}=\frac{4}{k^2}\Big( G^{-1}\fn{q^2=k^2}-G^{-1}\fn{q^2=0}\Big)\,.
\label{new def. of Planck mass}
}
This definition goes beyond the Einstein-Hilbert truncation since the contribution of higher-derivative invariants as the squared Weyl tensor can be included in $G^{-1}\fn{q^2=k^2}$. A similar definition was already used in Refs.\,\cite{Christiansen:2012rx,Christiansen:2014raa,Christiansen:2015rva,Meibohm:2015twa,Denz:2016qks,Christiansen:2017cxa,Eichhorn:2018akn,Eichhorn:2018ydy}, and the contributions of higher-derivative invariants were in particular investigated in Refs.\,\cite{Denz:2016qks,Eichhorn:2018ydy}. Since we treat the fixed-point value of $M_\text{p}^2\fn{k}/k^2$ as a free parameter, we may reinterpret our results as reflecting the definition \eqref{new def. of Planck mass}. We note that pure graviton vertices do not appear in the computation of the flow of the effective potential. They would differentiate between different terms in the gravitational sector. All this suggests that our estimate of the dominant graviton contribution, which is not affected by any gauge-fixing issues, is quantitatively rather robust.

Another robust result is the contribution from the measure sector (gauge fluctuations and ghosts) if physical gauge fixing is employed. The measure contribution results in a simple field-independent term for the flow of $U$. This term is necessary in order to account for the correct counting of physical degrees of freedom. The contributions from the matter sector are well understood as well. These are the well-tested standard contributions to the flow in flat space. One can extend their contribution to include Yukawa and gauge interactions with the Higgs field. As long as gauge couplings and Yukawa couplings remain small in the UV fixed-point region, these interaction effects on the flow of $U$ correspond to the standard perturbative beta function. The effect of the matter fluctuations on the field dependence of $U$ is much smaller than the gravitational contribution. As far as the critical exponents for $\tilde\lambda_H$ and $\tilde m_H^2$ are concerned, the interactions in the matter sector give only tiny corrections.

In the limit of a constant $\tilde M_{\text{p}*}^2$ and constant scalar wave function renormalization, the only term that is perhaps subject to somewhat larger truncation errors is the contribution from the physical scalar fluctuations of the metric $\tilde \pi_{0,g}$. Unless strong cancellations occur, this contribution is only a rather small fraction of the graviton, matter, and measure contributions. The uncertainty in the computation of this term will only result in a rather modest quantitative uncertainty for our overall results.

Replacing for the scaling form of the coefficient of the curvature scalar the constant $\tilde M_{\text{p}*}^2$ by a function $F\fn{\tilde \rho}$ of the scalar field induces additional terms in the flow equation for the mass term and quartic coupling, without affecting much the anomalous dimension $A$.
A small shift in the fixed-point value of $\tilde \lambda_H$ and in the location of the critical surface has no sizeable impact on the critical exponents. Finally, the omitted flow of the scalar wave function renormalization shifts $\theta_m$ from $2-A$ to $2-A-\eta_\phi$ and $\theta_\lambda$ from $-A$ to $-A-2\eta_\phi$. This would be a sizeable effect only if $\eta_\phi$ is of a similar magnitude as $A$.

We conclude that at least our Euclidean computation of the flow of the effective potential for the Higgs scalar seems rather reliable. A computation directly in Minkowski space along the lines discussed in Ref.\,\cite{Wetterich:2017ixo} would be welcome, but we do not expect important modifications as compared to the Euclidean results. The scalar potential is particularly robust with respect to analytic continuation since no momenta are involved. The remaining big issue is the determination of the fixed-point value of the dimensionless Planck mass $\tilde M_\text{p}^2$. This depends on $v_0$, rendering the flow of $\tilde M_\text{p}^2$ and $v_0$ a coupled system. Only once the fixed point of the combined system is found, it can be decided where a model is situated in the plane of Fig.\,\ref{fig:scalar number and Planck mass} or on the curves of Figs.\,\ref{fig:fixed point value of cosmological constant}--\ref{fig:critical exponent of quartic coupling}. Various computations give values of $\tilde M_\text{p}^2$. In the context of the present paper, the use of a physical gauge fixing is appropriate, and results from a gauge invariant formulation of the flow equation would be most welcome.

\subsection*{Acknowledgements}
We thank A.~Eichhorn, S.~Lippoldt, and M.~Schiffer for discussions.
This work is supported by ExtreMe Matter Institute (EMMI), the BMBF Grant 05P18VHFCA; 
by the Danish National Research Foundation under Grant No. DNRF:90; and by the DFG Collaborative Research Centre ``SFB 1225 (ISO- QUANT)."
M.\,Y.~is supported by the Alexander von Humboldt Foundation.

\begin{appendix}
\section{Effective action and formulation}\label{formulations}
In this Appendix we derive the flow equation for the effective scalar potential \eqref{flow equation of U} as an approximation to the exact flow equation for the effective average action. The exact flow~\cite{Wetterich:1992yh,Reuter:1993kw,Tetradis:1993ts,Ellwanger:1993mw,Morris:1993qb} takes a simple one-loop form:
\begin{align}
  \p_t \Gamma_k[\Phi] = \frac{1}{2}\text{Tr} \left[
    \left({\Gamma^{(2)}_k [\Phi] + {\mathcal R}_k } \right)^{-1}\, \p_t {\mathcal R}_k\right].
  \label{eq:wetterich}
\end{align}
Here, ${\mathcal R}_k$ is an infrared regulator function and $\p_t=k\p_k$. The trace in \eqref{eq:wetterich} sums over momenta and internal space indices of a multifield $\Phi$, and
the matrix of second functional derivatives $\Gamma^{(2)}_k$ is the full inverse propagator of
$\Phi$. For reviews, see~Refs.\,\cite{Berges:2000ew,Aoki:2000wm,Bagnuls:2000ae,%
  Polonyi:2001se,Pawlowski:2005xe,Gies:2006wv,Delamotte:2007pf,%
  Rosten:2010vm,Braun:2011pp}.
We apply this equation to a model of a singlet real scalar field and gravity and derive the flow equation for the effective scalar field.
Generalizations for additional fields are found in the main text.

\subsection{Setup}
We investigate an effective action of the type
\begin{align} 
  \Gamma_k=\Gamma_k^\text{gravity}+\Gamma_k^\text{Higgs}\,.
\end{align}
In the gravity sector we employ the Einstein-Hilbert truncation
\begin{align}
  \Gamma_k^\text{gravity} =
  -\frac{M_\text{p}^2}{2}\int\df^4x \sqrt{g}R +S_{\rm gf} +S_{\rm gh}
  \,,
			\label{effective action for gravity}
\end{align}
where $M_\text{p}$ is the reduced Planck mass related to Newton's
constant by $M_\text{p}^2=1/8\pi G_{\rm N}$.  The cosmological
constant is included in the scalar effective potential. 
The metric is linearly expanded around a fixed background
\al{ g_{\mu\nu}={\bar g}_{\mu\nu}+h_{\mu\nu}\,, } where
${\bar g}_{\mu\nu}$ is a constant background metric and $h_{\mu\nu}$
is a fluctuation field.  We will later use a flat Euclidean
background, ${\bar g}_{\mu\nu}=\delta_{\mu\nu}$.  
The gauge fixing and the ghost action for diffeomorphism symmetry are given by
\begin{align}\nonumber 
  S_{\rm gf} &=
                        \frac{1}{2\alpha}\int \df^4x\sqrt{\bar g}\,
                        {\bar g}^{\mu \nu}\Sigma_\mu\Sigma_{\nu} \,,
  \\[1ex]
  S_{\rm gh}	&=	-\int\df^4x\sqrt{\bar g}\,\bar C_\mu
               \left[ {\bar g}^{\mu\rho}{\bar \nabla}^2+
               \frac{1-\beta}{2}{\bar \nabla}^\mu{\bar \nabla}^{\rho}
               +{\bar R}^{\mu\rho}\right] C_{\rho}\,, \label{ghostaction}
\end{align}
where $C$ and $\bar C$ are ghost and antighost fields. 
A class of general gauge fixings is given by
\al{
\Sigma_\mu		
	= {\bar \nabla}^\nu h_{\nu \mu}-\frac{\beta +1}{4}{\bar \nabla}_\mu h\,,
	\label{gauge fixing function}
}
where $h={\bar g}^{\mu\nu}h_{\mu\nu}$ is the trace mode.
Bars denote covariant derivatives, etc., formed with the background metric.
Note that there are two gauge-fixing parameters for diffeomorphism symmetry,  $\alpha$ and $\beta$.
The parameter $\beta$ is dimensionless, whereas $\alpha$ has mass dimension minus 2.
For the physical gauge fixing, they are given by $\beta=-1$ and $\alpha\to 0$. We first keep general $\alpha$ and $\beta$ in order to see the particular role of the physical gauge fixing explicitly.

Next, we turn to the effective action for the Higgs sector.
In the SM, the Higgs field is a component of the doublet field, coupled to the $\text{SU}\fn{2}_L$ and $\text{U}\fn{1}_Y$ gauge fields as well as to quarks and leptons. 
Near the UV fixed point, the contributions from these couplings to the beta function are smaller than the ones of the graviton.
All essential points can be understood by restricting the discussion to a single real scalar field with $\mathbb{Z}_2$ symmetry as representing the physical mode of the Higgs boson.
The effective action takes the standard form
\al{
\Gamma_k^\text{Higgs}
	&=	\int\df^4x \sqrt{g}\Bigg[U\fn{\rho}
		+\frac{Z_\phi}{2} g^{\mu \nu}\,\p_\mu{\phi}\,\p_{\nu}\phi 
		\Bigg]\,.
				\label{effective action for matter}
}
We subsequently extend our findings to the SM or possible extensions.

The effective potential $U\fn{\rho}$ depends only on the invariant $\rho=\phi^2/2$.
The value of $U$ at the minimum can be identified with the cosmological constant.
We are interested in momenta much larger than the Fermi scale.
The expectation value of $\phi$ can be neglected in this range, and we expand
\al{
U=V+m_H^2\rho+\frac{1}{2}\lambda_H \rho^2+\cdots \,, 
\label{expansion of U}
}
where $m_H^2$ is the mass term of the Higgs boson and $\lambda_H$ is the quartic coupling.
The field-renormalization factor of $\phi$ is denoted by $Z_\phi$.

\subsection{Physical metric fluctuations}
A crucial quantity for the flow equation is the inverse propagator,
i.e.,\ the matrix of second functional derivatives of $\Gamma_k$.  The
physical understanding as well as calculational simplicity, are
greatly enhanced if we split the metric fluctuations into physical and
gauge fluctuations~\cite{Wetterich:2016vxu}.  In flat space, $\bar g_{\mu\nu}=\delta_{\mu\nu}$, one can use a momentum space representation. 

Let us start with splitting the metric fluctuations into
\al{
h_{\mu\nu}=f_{\mu\nu}+a_{\mu\nu}\,,
\label{metric decomposition}
}
where $f_{\mu\nu}$ are the physical metric fluctuations, which satisfy the transverse constraint $q^\mu f_{\mu\nu}=0$.  The physical metric
fluctuations can be decomposed into two independent fields as
\al{
f_{\mu\nu}=t_{\mu\nu}+s_{\mu\nu}\,,
\label{physical decomposition}
}
where the graviton $t_{\mu\nu}$ is the transverse and traceless (TT) tensor, i.e., $q^\mu t_{\mu\nu}=\delta^{\mu\nu}t_{\mu\nu}=0$.
The tensor $s_{\mu\nu}$ is given as a linear function of a scalar field $\sigma$ such that 
\al{
s_{\mu\nu}=\frac{1}{3}P_{\mu\nu} \sigma\,,
}
where we define the projection operator 
\al{
P_{\mu\nu} =\delta_{\mu\nu} -\frac{q_\mu q_\nu}{q^2}\,.
}

Similarly, the gauge modes or unphysical metric fluctuations $a_{\mu\nu}$ are decomposed into a transverse vector mode $\kappa_\mu$ satisfying $q^\mu \kappa_\mu=0$ and a scalar mode $u$. In summary, the metric fluctuations \eqref{metric decomposition} are parametrized by
\al{
f_{\mu\nu}&=t_{\mu\nu}+\frac{1}{3}P_{\mu\nu} \sigma\,, \notag \\
a_{\mu\nu}&=i(q_\mu \kappa_\nu+q_\nu \kappa_\mu) + \frac{q_\mu q_\nu }{q^2} u\,.
}
Using the linear combinations
\al{
\sigma&=\frac{3}{4}(h+q^2 s)\,,&
u&=\frac{1}{4}(h-3q^2 s)\,,&
}
we obtain the York decomposition~\cite{York:1973ia} of the fluctuation field
\begin{align}\nonumber 
h_{\mu\nu}&=t_{\mu\nu}+i(q_\mu \kappa_\nu  + q_\nu \kappa_\mu)\\[1ex]
&\quad
-\left( q_\mu q_\nu -\frac{1}{4}\eta_{\mu\nu}q^2 \right) s+\frac{1}{4}\eta_{\mu\nu}h,
\end{align}
where $h=\delta^{\mu\nu}h_{\mu\nu}$.
We see that the scalar modes $s$ and $h$ in the York decomposition are given as a mixture of the physical scalar mode $\sigma$ and the gauge mode $u$. 
The connection of the physical metric fluctuations $f_{\mu\nu}$ to the gauge invariant Bardeen potentials generally used in cosmology can be found in Ref.~\cite{Wetterich:2016vxu}. 

The decomposition yields Jacobians that read
\al{
J_\text{grav}=\left[ \det{}_{(1)}'(q^2) \right]^{1/2},~~
J_\text{gh}=\left[ \det{}_{(0)}''(q^2) \right]^{-1},
}
where a prime denotes a subtraction of the zero eigenmode.
These contributions are taken into account by introducing auxiliary fields
\al{ 
J_{\mathrm{grav}}&=\int 
{\mathcal D}\chi
{\mathcal D}\zeta{\mathcal D}\bar{\zeta}  \label{aux2}
 \\
&\quad\times\mathrm{exp}\Bigg\{-\int_q\,\,
\Big[ \frac{1}{2}\chi_\mu \left(q^2\right)^\prime \chi^\mu
-\bar{\zeta}_{\mu}\left(q^2\right)'\zeta^{\mu}
\Big]\Bigg\}\,, \notag
\\[1ex]
J_{\mathrm{gh}} &= \int {\mathcal D}\bar{\varphi}{\mathcal D}\varphi\,\mathrm{exp}\left[-\int_q\,\bar{\varphi}\,(q^2)'\varphi\right]\,,
\label{aux3}
}
where $\chi_\mu$ is a real bosonic vector field, $(\bar\varphi,\varphi)$ are complex bosonic scalar fields and $(\bar{\zeta}_\mu,\zeta^\mu)$ are vector anticommuting ghosts.

In flat space the matrix of second functional derivatives $\Gamma_k^{(2)}$ becomes block diagonal in the different representations of the Lorentz group, e.g., $t_{\mu\nu}$, $\kappa_\mu$ and the scalar fields ($\phi$, $\sigma$, $u$).
We will see that for the physical gauge, $\beta=-1$ and $\alpha\to 0$, it also becomes block diagonal in the physical fluctuations and gauge fluctuations.
Thus the scalar sectors decouple into separate sectors of $(\phi,\sigma)$ and the gauge mode $u$.
Finally, for a vanishing expectation value of $\phi$ the physical scalar sector becomes also block diagonal since $\sigma$ and $\phi$ belong to different representations of the $\mathbb{Z}_2$ symmetry.
This reflects the different representations of the Higgs doublet and the singlet contained in the metric.

For the $t_{\mu\nu}$-mode, we get
\al{
\left(\Gamma_{(tt)}^{(2)}\right)^{\mu\nu\rho\sigma}= \frac{M_\text{p}^2}{4}\left[ q^2-\frac{2U}{M_\text{p}^2} \right]P^{(t)\mu\nu\rho\sigma}\,,
}
where the TT-projection operator reads
\al{
P^{(t)\mu\nu\rho\sigma}=
\frac{1}{2}(P^{\mu\rho}P^{\nu\sigma}+P^{\mu\sigma}P^{\nu\sigma})-\frac{1}{3}P^{\mu\nu}P^{\rho\sigma}\,.
}
The Hessian for $\kappa_{\mu}$ is given by
\al{
\left(\Gamma_{(\kappa \kappa)}^{(2)}\right)^{\mu\nu}= \frac{1}{\alpha}q^2 \left[ q^2 -\alpha U\right] P^{(v)\mu\nu}\,,
}
with $P^{(v)}$ the projection operator on the vector mode, $P^{(v)}{}_\mu{}^\mu=3$.

In the $(\sigma,u,\phi)$-basis, the Hessian for the scalar modes becomes
\al{
\Gamma_{(00)}^{(2)}=
\pmat{
\begin{matrix} \left(\Gamma_{(00)}^{(2)}\right)_\text{grav} \end{matrix} & \begin{matrix}  \displaystyle \frac{U'\phi}{2}\\[10pt]  \displaystyle \frac{U'\phi}{2} \end{matrix} \\[15pt]
 \begin{matrix} \displaystyle \frac{U'\phi}{2} & & \displaystyle \frac{U' \phi}{2}\end{matrix}
&~~~~~
\displaystyle Z_\phi q^2+U'+2\rho U''
}\,,
\label{spin 0 matrix}
}
where the spin-0 gravitational part is given by the $2\times 2$ matrix
\begin{widetext}
\al{
\left(\Gamma_{(00)}^{(2)}\right)_\text{grav}=\pmat{
\displaystyle -\frac{M_\text{p}^2}{6} \left(q^2-\frac{U}{2M_\text{p}^2}\right)+\frac{(\beta +1)^2 }{16 \alpha } q^2
&&&&&
\displaystyle \frac{U}{4} + \frac{(\beta +1) (\beta -3)}{16 \alpha} q^2\\[20pt]
\displaystyle \frac{U}{4}+\frac{(\beta +1) (\beta -3)}{16 \alpha }q^2
&&&&&
\displaystyle -\frac{U}{4}+\frac{(\beta-3 )^2 }{16 \alpha } q^2
}\,.
\label{Hessian spin 0 part}
}
\end{widetext}
The choice of the gauge parameter $\beta=-1$ eliminates the off-diagonal terms in the matrix \eqref{Hessian spin 0 part}.
Furthermore, the $\sigma$-mode becomes independent of $\alpha$.
Thus the whole sector of physical metric fluctuations becomes independent of $\alpha$.
For the choice $\beta=-1$, the gauge fixing function \eqref{gauge fixing function} becomes $\Sigma_\mu={\bar \nabla}^\nu h_{\nu \mu}$.
Therefore, the choice $\beta=-1$ is a gauge fixing for which the gauge $\bar \nabla^\nu h_{\nu\mu}=0$ satisfies the transverse condition within the Faddeev-Popov method. 

Taking furthermore the limit $\alpha\to0$, and therefore realizing the physical gauge, the $u$-mode, i.e., the lower right element of the matrix \eqref{Hessian spin 0 part}, is dominated by $q^2/\alpha$.
In this limit, the finite part of this element (which involves $U$) no longer contributes after the inversion of $(\Gamma_k^{(2)}+{\mathcal R}_k)$.
The same thing holds for the mixing given as $U'\phi/2$ with the physical modes in the $3\times 3$ matrix \eqref{spin 0 matrix}.
For the physical gauge, one therefore deals with a decoupled gauge mode with inverse propagator $q^2/\alpha$ and two physical scalars with inverse propagator matrix
\al{
\left( \Gamma_{(00)}^{(2)}\right)_\text{ph}=\pmat{
\displaystyle -\frac{M_\text{p}^2}{6}\left(q^2 -\frac{U}{2M_\text{p}^2} \right) & U'\phi/2\\[15pt]
U'\phi /2& Z_\phi q^2+U' +2\rho U''
}.\label{physical modes matrix}
}
Finally, for $\alpha\to0$, the inverse propagator for the gauge-vector mode also becomes independent of $U$,
\al{
\left( \Gamma_k^{(2)}\right)_{\kappa\kappa}^{\mu\nu}=\frac{q^4}{\alpha}P^{(v)\mu\nu}.
}
We conclude that the total contribution of the gauge modes (vector + scalar) gives a contribution that is independent of $\phi$ and independent of all parameters in the effective action.

A similar property holds for the ghost contribution and the contribution from the Jacobians, \eqref{aux2} and \eqref{aux3}.
For an explicit computation, we decompose the ghost field as
\al{
C_\mu&=C_{\mu}^\perp+iq_\mu C\,,&
\bar C_\mu&=\bar C_{\mu}^\perp+iq_\mu \bar C\,,
}
where $C_{\mu}^\perp$ ($\bar C_{\mu}^\perp$) is the transverse (anti)ghost field and $C$ ($\bar C$) is the scalar (anti)ghost field.
The Hessians for the ghost fields is
\al{\nonumber 
\left(\Gamma_{(\bar C^\perp C^\perp)}^{(2)}\right)^{\mu\nu}&=-q^2P^{(v)\mu\nu}\,,\\[1ex] 
\Gamma_{(\bar CC)}^{(2)}&=\left( 2 -\frac{1+\beta}{2}\right)q^4\,.
}
We have demonstrated the decoupling of the gauge fluctuations for the physical gauge choice for the particular case of a flat background geometry.
This property holds actually for a general background geometry, as advocated in Ref.~\cite{Wetterich:2016ewc}.

\subsection{Flow generator from physical fluctuations}
Let us next investigate the structure of the flow equation in case of the decomposition presented in the previous subsection.
Since the sector of physical fluctuations decouples from the ones for the gauge modes we can treat their contributions separately, provided we choose a cutoff function that respects this decomposition.
This is achieved by a block-diagonal cutoff scheme with a physical cutoff $\mathcal R^\text{(ph)}_k$ that only acts on the fluctuations $f_{\mu\nu}$ and $\phi$ and a gauge cutoff $\sim \alpha^{-1}{\mathcal R}_k^{(g)}$ that only involves the gauge modes of the metric.
The contributions of the gauge fluctuations can be combined with the contributions of ghosts and Jacobians to a total measure contribution.
We write the general structure as
\al{
\p_t \Gamma_k=\zeta_k=\pi_k+\eta_k\,,
\label{beta function for the system}
}
with $\pi_k$ the physical mode contribution and $\eta_k$ the measure contribution.

For flat spacetime, we employ cutoff functions that replace for each mode $q^2$ by $P_k(q^2)=q^2+R_k(q^2)$.
This can be generalized by replacing $q^2$ with an appropriate covariant differential operator.
From the TT-mode $t_{\mu\nu}$, one finds a contribution
\al{
\pi_2=\frac{1}{2}{\rm Tr}_{(2)}\left. \frac{\p_t{\mathcal R}_k}{\Gamma _k^{(2)}+{\mathcal R}_k}\right|_{tt}
=\frac{5}{2}\int_q \frac{\p_t (M_\text{p}^2 R_k)}{M_\text{p}^2(P_k-k^2v)}\,.
\label{pi2 contribution in app}
}
Here, $\int_q=(2\pi)^{-4}\int \df^4 q$, and the factor $5$ comes from the trace of $P^{(t)}$, corresponding to the five independent degrees of freedom in $t_{\mu\nu}$.
The dimensionless quantity $v\fn{\rho}$ is defined as
\al{
v\fn{\rho}=\frac{2 U\fn{\rho}}{{M}_\text{p}^2k^2}\,.
\label{ratio of U and M}
}

The contribution of the two physical scalar fluctuations takes the form
\al{
\label{pi0 contribution in app}
\pi_0= \frac12 \int_q&{\tilde \p}_t \ln \bigg\{ 3\rho U'{}^2  \\
&+ \left(M_\text{p}^2P_k -\frac{U}{2} \right)\left( Z_\phi P_k+U'+2\rho U'' \right) \bigg\} \,, \notag
}
where $\tilde \p_t$ is the logarithmic derivative $k\p_k$ acting only on $M_\text{p}^2P_k$ and $Z_\phi P_k$, not on $U$ or $\rho$.
The mixing effects are small for the range of $\rho$ where
\al{
3\rho U'{}^2\ll \left(M_\text{p}^2P_k -\frac{U}{2} \right)\left( Z_\phi P_k+U'+2\rho U'' \right)\,.
\label{approximation for scalar propagator}
}
In this range, $\pi_0=\pi_{0,g}+\pi_{0,\phi}$ decouples into two separate parts.
The gravitational scalar contributes
\al{
\pi_{0,g}=\frac{1}{2}\int_q \frac{\p_t (M_\text{p}^2R_k)}{M_\text{p}^2(P_k-k^2v/4)}\,.
}
This contribution is similar to the tensor contribution $\pi_2$.
It is suppressed by a factor $1/5$, reflecting the single degree of freedom, and a smaller enhancement of the denominator for positive $v$.
The contribution of $\phi$,
\al{
\pi_{0,\phi}=\frac{1}{2}\int_q \frac{\p_t (Z_\phi R_k)}{Z_\phi P_k +U'+2\rho U''} \,,
}
is the standard expression for a real scalar theory.
For the full SM, it will be supplemented by contributions from the Goldstone directions, the gauge bosons, and fermions.

\subsection{Measure contribution}
We next turn to the measure contribution $\eta_k$.
We have already seen that for a physical gauge fixing it is independent of $\rho$ and $U$.
This contribution depends therefore only on the background metric.
We will establish that the total measure contribution takes for physical gauge fixing the simple overall form
\al{
\eta_k=-\frac{1}{2}{\rm Tr}_{(1)}\frac{\p_t P_k\fn{\mathcal D_1}}{P_k\fn{\mathcal D_1}}
-\frac{1}{2}{\rm Tr}_{(0)}\frac{\p_t P_k\fn{\mathcal D_0}}{P_k\fn{\mathcal D_0}}\,,
\label{contributions from eta}
}
with $\mathcal D_1$ and $\mathcal D_0$ appropriate differential operators formed with the background metric.
For a general metric they take the following forms:
\al{
{\mathcal D}_1&= -\bar \nabla^2-\frac{{\bar R}}{4}\,,&
{\mathcal D}_0&=-\bar \nabla^2-\frac{{\bar R}}{4}\,.&
}
Such a simple form has been proposed in Ref.~\cite{Wetterich:2016ewc}, based on a direct regularization of the Faddeev-Popov determinant.

Let us explicitly see that the contribution of the gauge modes and ghosts are given by \eqref{contributions from eta} in the present setup.
We first look at the contributions from the spin-1 gauge and ghost modes and the auxiliary field:
\al{\nonumber 
\eta_1&=\frac{1}{2}{\rm Tr}_{(1)}\left. \frac{\p_t{\mathcal R}_k}{\Gamma _k^{(2)}+{\mathcal R}_k}\right|_{\kappa\kappa}
-\left. \Tr_{(1)} \frac{\p_t {\mathcal R} _k}{\Gamma_k^{(2)}+{\mathcal R}_{k}}\right|_{\bar C^\perp C^\perp}\nn
&\quad 
+\frac{1}{2} \Tr_{(1)}\left. \frac{\partial_t {\mathcal R}_{k}}{ \Gamma^{(2)}_k + {\mathcal R}_{k}}\right|_{\chi\chi}
-\Tr_{(1)}\left. \frac{\p_t {\mathcal R}_{k}}{ \Gamma^{(2)}_k + {\mathcal R}_{k} }\right|_{\bar\zeta\zeta}.
\label{beta functions from spin 1}
}
The last two terms on the right-hand side are the contributions from the Jacobian \eqref{aux2} associated with the gauge mode $\kappa_\mu$.
For $\alpha\to 0$, the transverse vector metric fluctuation and the contributions from the Jacobian become
\al{
\delta_k^{(1)}&=\lim_{\alpha\to 0}\frac{1}{2}{\rm Tr}_{(1)}\left. \frac{\p_t{\mathcal R}_k}{\Gamma _k^{(2)}+{\mathcal R}_k}\right|_{\kappa\kappa}\nn
&\quad
+\frac{1}{2} \Tr_{(1)}\left. \frac{\partial_t {\mathcal R}_{k}}{ \Gamma^{(2)}_k + {\mathcal R}_{k}}\right|_{\chi\chi}
-\Tr_{(1)}\left. \frac{\p_t {\mathcal R}_{k}}{ \Gamma^{(2)}_k + {\mathcal R}_{k} }\right|_{\bar\zeta\zeta}\nn
&=\frac{1}{2}{\rm Tr}_{(1)}\,\frac{\p_t P_k}{P_k}\,.
\label{spin 0 xi contribution}
}
The contribution from the vector ghost mode takes the form
\al{
-\epsilon_k^{(1)}&=-\left. \Tr_{(1)} \frac{\p_t {\mathcal R} _k}{\Gamma_k^{(2)}+{\mathcal R}_{k}}\right|_{\bar C^\perp C^\perp}
=-{\rm Tr}_{(1)}\, \frac{\p_t P_k}{P_k}\,.
}

One finds a simple relation between the contributions from the gauge mode and the ghost field~\cite{Wetterich:2016ewc},
\al{
\epsilon_k^{(1)}=2\delta_k^{(1)}\,.
\label{relation between spin1 gauge and ghost modes}
}
The total contribution from spin-1 gauge modes is given by
\al{
\eta_1&=\delta_k^{(1)}-\epsilon_k^{(1)}=-\frac{1}{2}{\rm Tr}_{(1)}\, \frac{\p_t P_k}{P_k}\,.
\label{total spin 1 contribution}
}
A different normalization of the vector field does not change this result. If we redefine the transverse vector metric fluctuation as $\tilde\kappa_\mu=\sqrt{q^2}\kappa_\mu$, the contributions from the Jacobian in \eqref{beta functions from spin 1} are eliminated.
Instead, the contributions from the transverse vector metric fluctuation \eqref{spin 0 xi contribution} should be multiplied by a factor $1/2$. Consequently, the contributions from the spin-1 modes yield the result \eqref{total spin 1 contribution} independently of the field normalization.

Next, we discuss the contributions from the spin-0 modes involved in $\eta_k$.
We have
\al{
\eta_0&= \frac{1}{2} \Tr_{(0)}\left.\frac{\p_t {{\mathcal R}} _k}{\Gamma_k^{(2)}+{{\mathcal R}}_{k}}\right|_\text{gauge}\nn
&\quad
- \Tr_{(0)}\left.\frac{\p_t {{\mathcal R}} _k}{\Gamma_k^{(2)}+{{\mathcal R}}_{k}}\right|_{\bar C C}
+\Tr_{(0)}\left. \frac{\p_t {{\mathcal R}}_{k} }{ \Gamma^{(2)}_k + {{\mathcal R}}_{k}}\right|_{\bar{\varphi}\,\varphi} .
}
The last term on the right-hand side corresponds to the contribution from the auxiliary fields for the Jacobian \eqref{aux3} associated with the spin-0 ghost mode.
Here, we denote the first term on the right-hand side corresponding to the gauge scalar mode of the metric fluctuation as
\al{
 \delta_k^{(0)}&=\lim_{\alpha\to 0}\frac{1}{2} \Tr_{(0)}\left.\frac{\p_t {{\mathcal R}} _k}{\Gamma_k^{(2)}+{{\mathcal R}}_{k}}\right|_\text{gauge}\nn
&=\frac{1}{2}\Tr_{(0)}\,\frac{\p_t P_k}{P_k}\,.
\label{contribution from gauge mode of graviton}
}
The spin-0 ghost and the auxiliary fields give 
\al{
-\epsilon_k^{(0)}&=- \Tr_{(0)}\left.\frac{\p_t {{\mathcal R}} _k}{\Gamma_k^{(2)}+{{\mathcal R}}_{k}}\right|_{\bar C C}
+\Tr_{(0)}\left. \frac{\p_t {{\mathcal R}}_{k} }{ \Gamma^{(2)}_k + {{\mathcal R}}_{k}}\right|_{\bar{\varphi}\,\varphi} \nn
&=-\Tr_{(0)}\,\frac{\p_t P_k}{P_k}\,.
}
Within the spin-0 modes, a relation similar to \eqref{relation between spin1 gauge and ghost modes} holds, i.e.,
\al{
\epsilon_k^{(0)}=2\delta_k^{(0)}\,.
\label{relation between spin0 gauge and ghost modes}
}
The total spin-0 contribution from the gauge and the ghost modes is
\al{
\eta_0=\delta_k^{(0)}-\epsilon_k^{(0)}=-\frac{1}{2}\Tr_{(0)}\,\frac{\p_t P_k}{P_k}\,.
}
Again, we may redefine the gauge scalar mode as $\tilde u=u/\sqrt{q^2}$; the contribution from the gauge scalar mode \eqref{contribution from gauge mode of graviton} has to be multiplied by the factor 2. On the other hand, the contributions from the Jacobian corresponding to the gauge scalar mode are modified correspondingly. Therefore, the total contributions from the gauge scalar mode and the contributions from the Jacobian do not change from \eqref{contribution from gauge mode of graviton}, and accordingly the relation \eqref{relation between spin1 gauge and ghost modes} holds.

In a flat background, the contributions from the gauge modes are given by
\al{
\eta_k&=\eta_1+\eta_0\,,
}
with
\al{
\eta_1&=-\frac{3}{2}\int_q \frac{\p_t P_k\fn{q^2}}{P_k\fn{q^2}}\,,&
\eta_0&=-\frac{1}{2}\int_q\frac{\p_t P_k\fn{q^2}}{P_k\fn{q^2}}\,,&
\label{eta contributions in app}
}
since the differential operators $\mathcal D_1$ and $\mathcal D_0$ are simply given by $q^2$ multiplied with appropriate projectors.

\subsection{Flow of the scalar potential}
To summarize, the flow generator \eqref{beta function for the system} consists of the four components 
\al{
\zeta_k=\pi_2+\pi_{0}+\eta_1+\eta_0\,.
\label{total contribution in the system}
}
With the approximation \eqref{approximation for scalar propagator}, we have $\pi_0=\pi_{0,g}+\pi_{0,\phi}$.
In the high-momentum range and for constant $M_\text{p}^2$ and $Z_\phi$ and small $v$, all fluctuations behave as for massless particles, with contributions $\sim \int_q(\p_t P_k)/P_k$. One can easily count the degrees of freedom: The physical modes ($\pi_2+\pi_{0,g}+\pi_{0,\phi}$) have $7=5+1+1$ degrees of freedom. From these degrees are subtracted the $4=3+1$ degrees of freedom from ($\eta_1+\eta_0$). The remaining $7-4=3$ degrees of freedom correspond to the three propagating modes of the system, namely the two helicities from the graviton and 1 degree of freedom from the real scalar.

We  finally evaluate the explicit form of the flow generator \eqref{beta function for the system} and \eqref{total contribution in the system} for a flat background and consider a constant scalar field. 
Dividing out a total volume factor, $\zeta_k$ generates directly the flow of the effective potential $U$,
\al{
\p_t U=\tilde \pi_2+\tilde\pi_0+\tilde \eta_1+\tilde \eta_0\,.
}
We employ the Litim-type cutoff function~\cite{Litim:2001up} for the regulator
\al{
R_k=(k^2-p^2)\theta\fn{k^2-p^2}\,.
}
With this regulator, one can perform the momentum integrations analytically and obtain the explicit form of the beta functions.
Verifying that the $\delta$ function in $\p_t R_k$ does not contribute, we can use $\p_t P_k=\p_t R_k=2k^2 \theta\fn{k^2-p^2}$. The momentum integrations of Eqs.\,\eqref{pi2 contribution in app}, \eqref{pi0 contribution in app}, and \eqref{eta contributions in app} yields the flow equation \eqref{pi0 contribution} for the effective potential, with $\tilde \eta_k=\tilde \eta_0+\tilde \eta_1$.
The measure contributions are simply given by
\al{
\tilde\eta_1&=-\frac{3k^4}{32\pi^2}\,,&
\tilde\eta_0&=-\frac{k^4}{32\pi^2}\,.&
}

\end{appendix}

\bibliography{refs}
\end{document}